\documentclass[iop]{emulateapj}
\usepackage{amsmath, graphicx, apjfonts, gensymb, hyperref, natbib}
\def\apj{{ApJ}}                 
\def\apjl{{ApJ}}                
\def\apjs{{ApJS}}               
\def\mnras{{MNRAS}}             
\def\prd{{Phys.~Rev.~D}}        
\def\nat{{Nature}}              

\def\jcap{{JCAP}}               

\def\physrep{{Phys.~Rep.}}   

\newcommand{\vect}[1]{\mathbf{#1}}
\newcommand{\msun}{h^{-1} M_{\odot}}
\newcommand{\kms}{{\rm km}\, {\rm s}^{-1}}

\shorttitle{Bulk flow in $\Lambda$CDM simulation}

\begin{document}

\title{Bulk flow of halos in $\Lambda$CDM simulation}

\author{Ming Li\altaffilmark{1}, 
Jun Pan\altaffilmark{1,2}, 
Liang Gao\altaffilmark{2}, 
Yipeng Jing\altaffilmark{3}, 
Xiaohu Yang\altaffilmark{3},
Xuebin Chi\altaffilmark{4}, 
Longlong Feng\altaffilmark{1},
Xi Kang\altaffilmark{1,5}, 
Weipeng Lin\altaffilmark{3},
Guihua Shan\altaffilmark{4}, 
Long Wang\altaffilmark{4}, 
Donghai Zhao\altaffilmark{3}, 
Pengjie Zhang\altaffilmark{3}}
\altaffiltext{1}{Purple Mountain Observatory, 2 West Beijing Rd., 
Nanjing 210008, P. R. China}
\altaffiltext{2}{National Astronomical Observatories, 
Chinese Academy of Sciences, 20A Datun Rd., Chaoyang District, Beijing 100012, 
P. R. China}
\altaffiltext{3}{Key Laboratory for Research in Galaxies and Cosmology, 
Shanghai Astronomical Observatory, 80 Nandan Rd., Shanghai 200030, P. R. China}
\altaffiltext{4}{Supercomputing center, Computer Network Information Center, Chinese
Academy of Sciences, 4 Zhongguancun Nansijie, Haidian District, Beijing 100190, China}
\altaffiltext{5}{Partner Group of MPI for Astronomy, PMO, 2 West Beijing Road, Nanjing 210008, China}

\email{jpan@bao.ac.cn}
\begin{abstract}
Analysis of the Pangu $N$-body simulation
validates that the bulk flow of halos follows a Maxwellian distribution which variance
is consistent with the prediction of the linear theory of structure formation. We propose
that the consistency between the observed bulk velocity and theories
should be examined at the effective scale of the radius of a spherical top-hat window
function yielding the same smoothed velocity variance in linear theory as the sample window function does.
We compared some recently estimated bulk flows from observational
samples with the prediction of the $\Lambda$CDM model we used; some results
deviate from expectation at a level of $\sim 3\sigma$ but the discrepancy is
not as severe as previously claimed. We show that bulk flow is only weakly correlated with the
dipole of the internal mass distribution, the alignment angle between the mass dipole
and the bulk flow has a broad distribution peaked at $\sim 30-50^\circ$, and also that the bulk flow
shows little dependence on the mass of the halos used in the estimation.
In a simulation of box size $1h^{-1}$Gpc, for a cell of radius $100^{-1}$Mpc the 
maximal bulk velocity is $>500\kms$,
dipoles of the environmental mass outside the cell are not tightly aligned with the bulk flow, 
but are rather located randomly around it with separation angles  $\sim 20^\circ$--$40^\circ$. In the fastest
cell there is a slightly smaller number of low-mass 
halos; however halos inside are clustered more strongly at scales  $\gtrsim 20h^{-1}$Mpc, 
which might be a significant feature since the correlation between bulk flow and halo 
clustering actually increases in significance beyond such scales.
\end{abstract}


\section{Introduction}
Bulk flow refers to the apparent coherent peculiar motion of galaxies and galaxy clusters in
a considerably large volume around us. In practice there are several ways to 
estimate bulk flows from various observation resources, such as galaxy catalogs from 
peculiar velocity surveys \citep[e.g.][]{FeldmanEtal2010},  
compiled Type Ia supernovae data \citep[e.g.][]{DaiEtal2011},  and galaxy clusters in combination with
cosmic microwave background (CMB) observations \citep[e.g.,][]{KashlinskyEtal2010}. 
Recently some interesting new methods based on galaxy two-point correlation functions 
\citep{SongEtal2011} and galaxy light \citep{NusserEtal2011, AbateFeldman2012}
have also emerged. 

Analysis of  the spiral galaxy catalog of the SFI++ survey \citep{SpringobEtal2007}
shows that within a top-hat spherical window 
of radius $40h^{-1}$Mpc the velocity of the bulk flow is $338\pm 38\kms$ toward Galactic plane
$(l,b)=(276^\circ,14^\circ)$ 
with a $3^\circ$ $1\sigma$ uncertainty, and then $257\pm 44\kms$  toward $(l,b)=(279^\circ, 10^\circ)$ 
with a $6^\circ$ error within window of radius $100h^{-1}$Mpc \citep{NusserDavis2011}. 
These measurements are in agreement with the analysis by \citet{SandageEtal2010}  of data consisting of 
supernovae, selected nearby galaxies, and galaxy clusters.  \citet{FeldmanEtal2010} constructed a 
composite catalog of galaxies with
peculiar velocities measured in different surveys, including the SFI++. They estimate that the bulk flow
within a Gaussian window of $50h^{-1}$Mpc is $416\pm 78\kms$ in the direction 
$(l,b)=(282\pm11^\circ, 6\pm6^\circ)$ \citep[see also][]{WatkinsEtal2009}.

Employing the peculiar velocities of supernovae is another viable route to detect bulk flow, though
such samples are usually very sparse and prone to Malmquist bias. \citet{DaiEtal2011} fitted a 
bulk flow of $188^{+119}_{-103}\kms$  in the direction $(l,b)=({290^{+39}_{-31}}^\circ, {20^{+32}_{-32}}^\circ)$
to the Union2 supernovae catalogue \citep{AmanullahEtal2010} for redshifts $z<0.05$, but no significant
bulk flow was detected from data at $z>0.05$.  \citet{ColinEtal2011} used the same data to obtain a
similar estimate but with a higher median amplitude of $250^{+190}_{-160}\kms$. 
However, using a different supernovae data set within the redshift shell  $z=(0.0043, 0.028)$, 
\citet{WeyantEtal2011} estimate that the local flow is
$538\pm86\kms$ pointing to $(l,b)=(258^\circ\pm 10^\circ, 36^\circ\pm11^\circ)$, or $446\pm101\kms$ towards 
$(l,b)=(273\pm11^\circ, 46\pm 8^\circ)$ if a different technique is employed, this is in agreement with
the dipole of the CMB $(l,b)=(263^\circ.99\pm0^\circ.14, 48^\circ.26\pm0^\circ.03)$ \citep{JarosikEtal2011}.  
\citet{JhaEtal2007} and \citet{HaugbolleEtal2007} have found the same values with similar uncertainties.

The availability of recent galaxy peculiar velocity data is limited to our local universe; the bulk flow at 
higher redshift, sometimes dubbed {\em dark flow}, is mainly explored through the
kinetic Sunyaev--Zel'dovich (kSZ) effect of galaxy 
clusters \citep{SunyaevZeldovich1980}. \citet{KashlinskyEtal2011} computed kSZ 
signals of 771
X-ray clusters in the 7 yr {\em Wilkinson Microwave Anisotropy Probe} (WMAP) 
CMB map, and conclude that the flow at $z\leq 0.16$ is directed to
$(l,b)=(278\pm18^\circ, 2.5\pm15^\circ)$ and then
$(l,b)=(283\pm19^\circ,20\pm15^\circ)$ if $z\leq 0.25$. They further argue that the flow
at these depths shall reach magnitude of $\sim1000\kms$ according to an earlier investigation in
\citet{KashlinskyEtal2010}. \citet{OsborneEtal2011} derived a conflicting assertion
from the same CMB map in conjunction with 736 {\em ROSAT} observed clusters: that
there is no significant detection of kSZ effects at low multipoles, basically denying the existence of
bulk flow. In some cases, however, the thermal Sunyaev-Zel'dovich effect might
induce a dipole that could easily be misunderstood as bulk flow of $\sim 2000-4000\kms$.

Bulk flow is a topic of long-term interest to observational  cosmology \citep[see][ for a review of 
early works]{StraussWillick1995},  and special surveys have been
dedicated to it \citep[e.g.][]{CourtoisEtal2011}. 
However, as we see, no consensus on the amplitudes, directions, or convergence depth of bulk flows has yet been
achieved to reconcile different measurements.
Nonetheless, some authors have argued that the amplitude of their measured bulk flow
is too strong over such large scales, presenting a challenge to the standard $\Lambda$CDM
model, or at least to that of the 5yr WMAP parameters 
\citep{WatkinsEtal2009,  KashlinskyEtal2010, FeldmanEtal2010, MacaulayEtal2011}. Not surprisingly, 
cosmological models of different flavors have been constructed to explain such anomalies
\citep[e.g.,][]{MersiniHolman2009, AfshordiEtal2009, WymanKhoury2010}, 
but new analysis of similar data sets seems to have nullified support 
for such a violation \citep{TurnbullEtal2012, MaScott2012}.

Expectation about bulk flow in  $\Lambda$CDM universe is generally calculated with
linear perturbation theory of large scale structure in that at large scales its
accuracy is believed sufficient. Often the 1-D velocity variance of dark matter
are quoted to compare with measurements, however we need to address here that it is
the rms velocity that should be used instead of the 1-D rms velocity. \citep{MakEtal2011}
calculated that by linear theory the rms bulk velocity is typically $\sim 300\kms$, and claimed that 
uncertainty at $95\%$ confidence due to sample variance is
approximately $200\kms$ for a top-hat window of radius $60h^{-1}$Mpc \citep{MakEtal2011}.
A concern is that non-linearity might not be negligible even at very 
large scales \citep[e.g.][]{Scoccimarro2004}, which could act as systematical bias to the 
conclusion about the consistency between model and data. 

Although linear theory can predict the possibility of observing a bulk flow of particular 
amplitude at certain scale,
several key problems yet can not be easily tackled analytically,  e.g. internal properties of the volume 
demonstrating large bulk blow. 
Practically halo catalogues from N-body simulation in large box 
with sufficient mass resolution are best suited for such task.
The reason of focusing on halos instead of dark matter is that observational objects used
to determine bulk flow are galaxies and galaxy clusters which are residing 
in halos, and in practice the strongly non-linearity of peculiar velocities of 
galaxies is largely filtered out so that
what contribute to bulk flow estimation is mainly the motion following their 
host halos \citep[e.g.][]{WatkinsEtal2009}. 
In fact \citet{BahcallEtal1994} and \citet{MoscardiniEtal1996} have performed
analysis of mock halo catalogues and obtained useful results,  but their simulations are 
either of very low mass resolution or based on compromised simulation method.
In this paper we will demonstrate our analysis of the velocity 
field of halos resolved from a dark matter only $\Lambda$CDM 
simulation in $1h^{-1}$Gpc box with $3072^3$ particles. The large volume
and high mass resolution of our simulation enables investigating halo
behaviors in detail over broad dynamic ranges superseding previous works.

In section 2 definition and basic theoretical prediction of bulk flow is introduced, which is followed 
by section 3 presenting measurements of bulk flow of randomly placed cells in simulation. 
Section 4 is devoted to analysis of special regions showing extraordinarily large bulk flow velocity.
Summary and discussion is in the last section.

\section{Bulk flow in linear theory}
Placing  a window of characteristic scale $R$ randomly in the sample space,  
if $N$ objects
(galaxies, galaxy clusters or halos, in this work just the latter) are 
enclosed, bulk flow of 
the particular volume indicated by the particular object is
\begin{equation}
\vect{V}=\sum_{i=1}^N w_i \vect{v}_i/\sum w_i\ ,
\label{eq:vb}
\end{equation}
in which $\vect{v}_i$ is the peculiar velocity of the $i$th object and $w_i$ is 
the weight assigned.
Practically the weights could be originated from 
radial selection function, 
angular selection function (survey mask), luminosity, mass and etc.

\subsection{Sample window}
Effect of the sample window is pure geometrical, which can be easily modeled, 
In continuous limit, Eq.~\ref{eq:vb} becomes
\begin{equation}
\vect{V}=\vect{v}\otimes W(R) = \int \vect{v}(\vect{r}) W(\vect{r}; R) {\rm d} r^3=\frac{1}{(2\pi)^3}\int \vect{v}(\vect{k}) 
\widetilde{W} {\rm d}^3k\ ,
\label{eq:vdef}
\end{equation}
where $W(\vect{r};R)$ is the window function of characteristic scale $R$ evaluated at position vector $\vect{r}$
and $\widetilde{W}$ is its Fourier transformation. In principle $W$ could be anisotropic, e.g. due
to incomplete sky coverage and non-uniform depth. 
The simplest and mostly common are the spherical top-hat 
window $\widetilde{W}_{th}=3(\sin kR - kR \cos kR)/(kR)^3$ and Gaussian
window $\widetilde{W}_G=\exp(-k^2R^2/2)$, in fact there are little differences among
top-hat, Gaussian and anisotropic windows in for bulk flow statistics if effective scale 
has been taken care of. 
Sometimes bulk flow estimation is provided by objects within a spherical shell defined 
by two radius $R_1 >R_0$, 
it is easy to see that Eq.~\ref{eq:vdef} applies with window function 
$\widetilde{W}_S=(R_1^3 \widetilde{W}_1-R_0^3 \widetilde{W}_0)/(R_1^3-R_0^3)$.

Probability distribution function (PDF) of $\vect{V}$ could be expressed as 
\begin{equation}
p(\vect{V}){\rm d} \vect{V}=p(V) {\rm d} V p(\hat{n}_V) {\rm d} \hat{n}_V
\label{eq:Vpdf}
\end{equation}
where $\hat{n}_V$ denotes the unit vector in direction of the bulk velocity. 
Isotropic assumption
leads to $p(\hat{n}_V)=1/4\pi$ and $\langle \hat{n}_V \rangle=0$, which ensures that
$\langle \vect{V} \rangle=\int V p(V){\rm d}V \int \hat{n}_V p(\hat{n}_V){\rm d} \hat{n}_V
=\langle V \rangle \cdot  \langle \hat{n}_V \rangle=0$. Integration over the 
angular part of
Eq.~\ref{eq:Vpdf} yields the PDF of the amplitude of bulk flow
\begin{equation}
p(V){\rm d}V=\left[ \int p(\vect{V}) {\rm d}\hat{n}_V \right] 
V^2{\rm d}V=4\pi p(\vect{V})V^2{\rm d}V\ .
\end{equation}
$\vect{V}$ by definition is the velocity field smoothed by the window function, once the
smoothing scale is sufficiently large the distribution of $\vect{V}$ shall be 
very close to Gaussian so that Maxwellian distribution could be invoked
to model $p(V)$ \citep{BahcallEtal1994}
\begin{equation}
p(V)dV=\sqrt{\frac{2}{\pi}}  \left( \frac{3}{\sigma_V^2} \right)^{3/2} V^2 \exp\left( -\frac{3 V^2}{2\sigma_V^2} \right) {\rm d} V\ , 
\label{eq:linVpdf}
\end{equation}
where the variance of $V$ can be obtained with $P_{\vect{v}\vect{v}}$, the power spectrum of $\vect{v}$, 
through
\begin{equation}
\sigma_V^2=\frac{1}{(2\pi)^3}\int P_{\vect{v}\vect{v}} {\widetilde{W}}^2 {\rm d}^3k \ .
\end{equation} 
Given that radius of the window function $W(R)$ deployed to measure bulk flow is 
fairly large, one would  expect 
that the velocity field smoothed at such scale will be well described by
linear evolution of  the initial condition, then if the initial distribution of velocity 
is Gaussian, for instance in
the case of our simulation, Eq.~\ref{eq:linVpdf} shall be a good approximation to the 
PDF of amplitudes of bulk flows.
With the model, the most likely amplitude of bulk flow is simply $V_p=\sqrt{2/3} \sigma_V$, which
ranges of variance corresponding to different levels can be computed by the integral
\begin{equation}
\int p(V) {\rm d}V = {\rm erf} \left( \sqrt{ \frac{3V^2}{2\sigma_V^2}} \right) - \sqrt{\frac{6V^2}{\pi\sigma_V^2}} 
\exp\left(-\frac{3V^2}{2\sigma_V^2}\right) \ .
\label{eq:range}
\end{equation}

If $\vect{v}$ is curl free, a velocity potential field can be defined as 
$\theta(\vect{r})=-\nabla \cdot \vect{v}/(Haf)$ with the scale factor
$a=1/(1+z)$,
$f\equiv {\rm d}\log D(a)/{\rm d}\log a\approx \Omega_m^{4/7}+(1+\Omega_m/2)\Omega_\Lambda/70$ 
and $D(a)$ is the linear density growth factor at redshift $z=1/a-1$
\citep{LahavEtal1991} \footnote{A better approximation to $f$ can be found in \citet{Linder2005}.}, we have
\begin{equation}
\sigma^2_V=\frac{(Haf)^2}{(2\pi)^3}  \int  \frac{P_{\theta\theta}}{k^2}\widetilde{W}^2 {\rm d}^3 k\ .
\label{eq:sigV}
\end{equation}
The above expression relies on the assumption of negligible rotational velocity. In the linear regime,
if the biasing of halo velocity to dark matter velocity is unity, it
can be further simplified with the approximation $P_{\theta\theta}\approx P_{\delta\delta}^{(L)}$ where 
$P_{\delta\delta}^{(L)}$ is the linear matter power spectrum.

\subsection{Selection function}
In reality large fraction of galaxy samples are magnitude-limited (or flux-limited), 
galaxies fainter than certain threshold are missed in the sample, so that
number density of observed galaxies $n(r)$ as function of distance to the observer,
termed as radial selection function,
is not constant. In the presence of selection function, if no correction is made, the
measured bulkfow local to an observer is 
\begin{equation}
\vect{V}=\frac{\int \vect{v}(\vect{r}) n(r) W(\vect{r}; R) {\rm d}^3 r}{\int n(r) W(\vect{r};R) {\rm d}^3 r}\ ,
\label{eq:Vs}
\end{equation}
in which $n(r)$ acts as the weighting function. In theoretical modeling Eq.~\ref{eq:Vs}
is equivalent to Eq.~\ref{eq:vdef} if a new window function
is defined through $W_O=n(r)W/\int nW {\rm d}^3 r$, however  there is still the conceptual
difference of applying selection function than a pure geometrical window function.
A non-constant selection function reflects the fact that the sampling to the velocity field is
distance dependent; a window function not of top-hat type rather simply denotes that
in the estimation the velocity field
is weighted by a particular scheme, but the sampling to the field is fair.

The simplest proposal to correct the unfair sampling rate depicted by the selection
function is to divide the measured peculiar velocity of an object by the selection function, 
\begin{equation}
\hat{\vect{V}}=\frac{\sum_i^N w_i\vect{v}_i /n(r_i)}{\sum_i^N w_i/n(r_i)}\ ,
\label{eq:sel}
\end{equation}
we will check its performance numerically later in this report (subsection \ref{subsec:sel}).

Note that the discussion here is also applicable to the angular selection function which is 
termed as the completeness mask defined as the ratio of number of observed objects
to the local observable number of that type of objects.

\subsection{Physical Weights}
Another type of weights different to selection functions is coming from physical properties
of astronomical objects, such as mass, luminosity, internal velocity dispersion and etc. This 
kind of weights can not be assimilated into the sample geometrical window function in
theoretical works, instead we have to develop statistical models to account for effects
of these weighting schemes which often involves calculation of series of correlation
between peculiar velocity and object's physical quantities. 
Among the various physical weights, 
probably the most commonly seen is the mass. 
Mass may not always be the dominant actor determining properties of galaxies and clusters, 
but is always a major facotr. For example, for galaxies in some bands their luminosity-mass relation is
considerably tight,
weighting by luminosity could be deemed roughly equivalent to the weighting by 
certain power of mass. So later in this paper we will back to the issue of bulk flow
weighted by mass, with a demonstrative numerical analysis (subsection \ref{subsec:mv}).

\section{Bulk flow shown in the Pangu simulation}
\subsection{The Pangu simulation and its halo catalogue}
The Pangu simulation (PS-I) is a large volume and high resolution simulation, 
carried out under the scheme of
the Computational Cosmology Consortium of China (dubbed C4). 
PS-I assumes a $\Lambda$ cold dark matter ($\Lambda$CDM) cosmology model with parameters
\begin{eqnarray}
  \label{eq:cosmo_params}
  & & \Omega_m = \!0.26, \; \Omega_b=0.044, \; \Omega_{\Lambda}=0.74, \nonumber \\
  & & h = 0.71, \; \sigma_8=0.8, \; n_s=1 \nonumber \ .
\end{eqnarray}
The simulation contains dark matter only, and uses $N=3,072^3$ particles to follow 
the distribution 
and evolution of dark matter within a periodic box with $L=1000 h^{-1}$Mpc on a side. 
Each particle has 
a mass of $2.48915 \times 10^9 \msun$. The Plummer-equivalent force softening 
length is kept constant of $7 h^{-1}$kpc.

The PS-I starts from redshift $z_{init}=127$, initial positions and velocities of 
particles is generated 
with Zel'dovich approximation from a glass-like particle set. Input linear power 
spectrum is computed
with the {\tt CAMB} \citep{LewisEtal2000}. The simulation is then run 
with {\tt L-GADGET}, a 
memory-optimized version of {\tt GADGET2} \citep{Springel2005}. 
{\tt L-GADGET2} is designed to meet requirements of high performance 
computations, only the 
tree-particle mesh algorithm is included to calculate the 
gravitational forces efficiently. 
Totally 64 snapshots are saved from $z_{init}=127$ to redshift $z=0$. 
The PS-I is performed on 
the supercomputer {\tt Lenovo Deepcomp7000} at 
Supercomputing Center of Chinese Academy of Sciences. We use 2048 cores and 
about 6TB memory 
at the peak time. The simulation consumes approximately 
$6.5 \times 10^5$ CPU hours (about 13 
days) in total and consists of 6151 time steps.

Dark matter halos are identified on the fly during the simulation for 
each snapshot, using the 
standard friends-of-friends (FOF) algorithm with linking length of 0.2 
times the mean particle separation. 
Each FOF group must contain at least 20 particles, at redshift $z=0$ 
there are $4.2 \times 10^7$ identified particle groups. 
There are a little bit of ambiguity in definition of halo mass. Halo mass 
mostly used in literature is defined as the 
mass enclosed by a sphere centered on the halo center with certain radius, 
within which the average density 
is some factor larger than the critical density. A more convenient definition 
of halo mass is the total mass 
of all member particles of the FoF group, which is used in this paper. 
For simplicity the average of velocities of member particles are taken as the velocity of 
the halo. 

In this work we used a subset of the group catalogue as our full halo catalogue, 
which consists of FoF groups of mass  larger than $2.5\times 10^{11}\msun$ only and 
makes  $1.28\times 10^7$ entries in the end.
Selected FoF groups contains at least 100 member particles, corresponding 
Poisson fluctuation is greater than $10$ which is the threshold used for FoF 
group identification, limiting
discreteness error to $10\%$ level. Another reason is that it is not easy to
detect large number of low mass halos in observation. Our prudence test verifies
that including FoF groups with number of particle $<100$
does not make any significant amendment to final results, even though effects of non-linearity are
stronger. More importantly we will show that statistics of
 bulk flow are not sensitive to the mass of halos in the sample (subsection~\ref{subsec:sub}).

\subsection{The probability distribution function of bulk flow}

\begin{figure*}
\resizebox{\hsize}{!}{\includegraphics{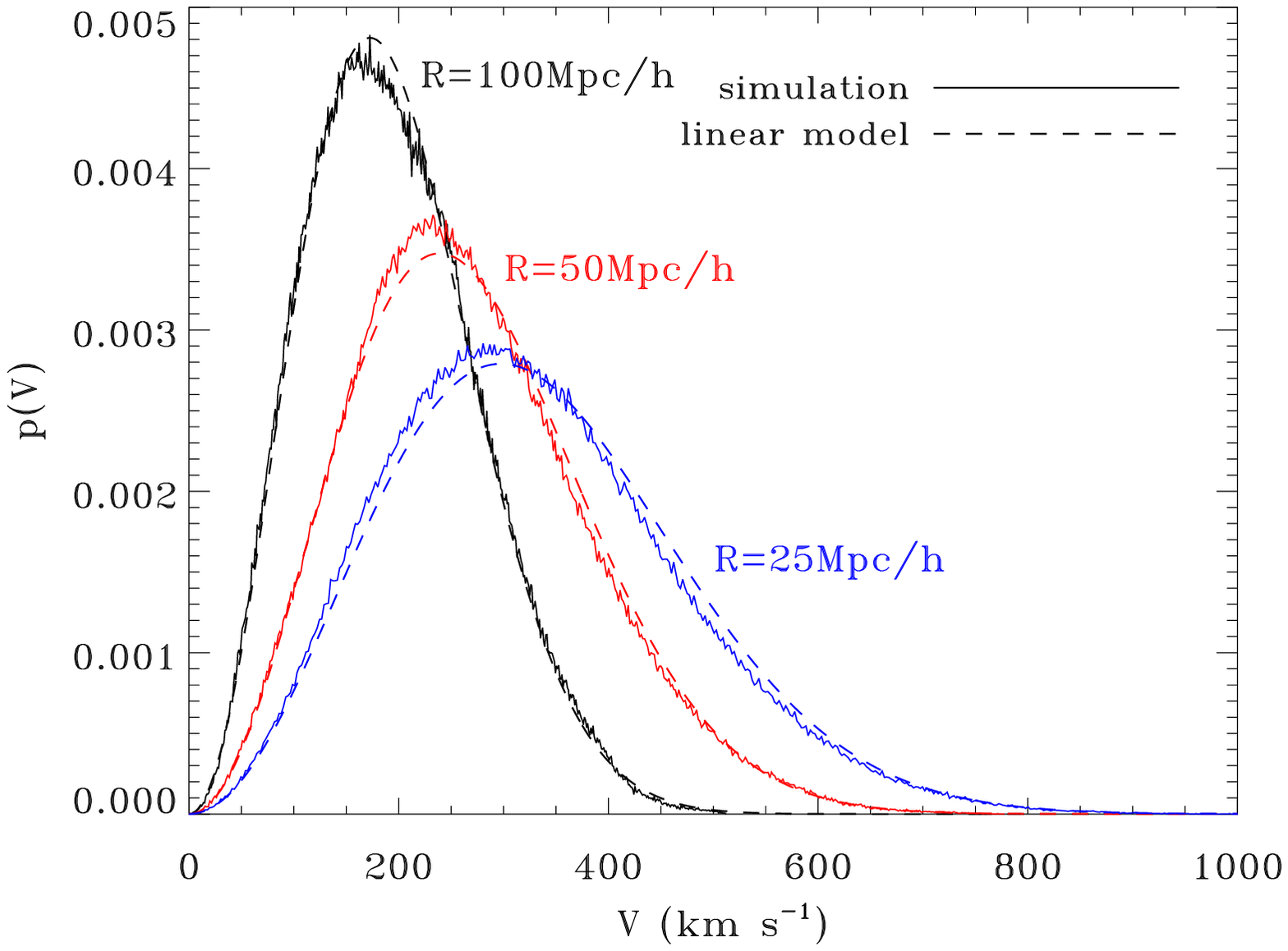}\includegraphics{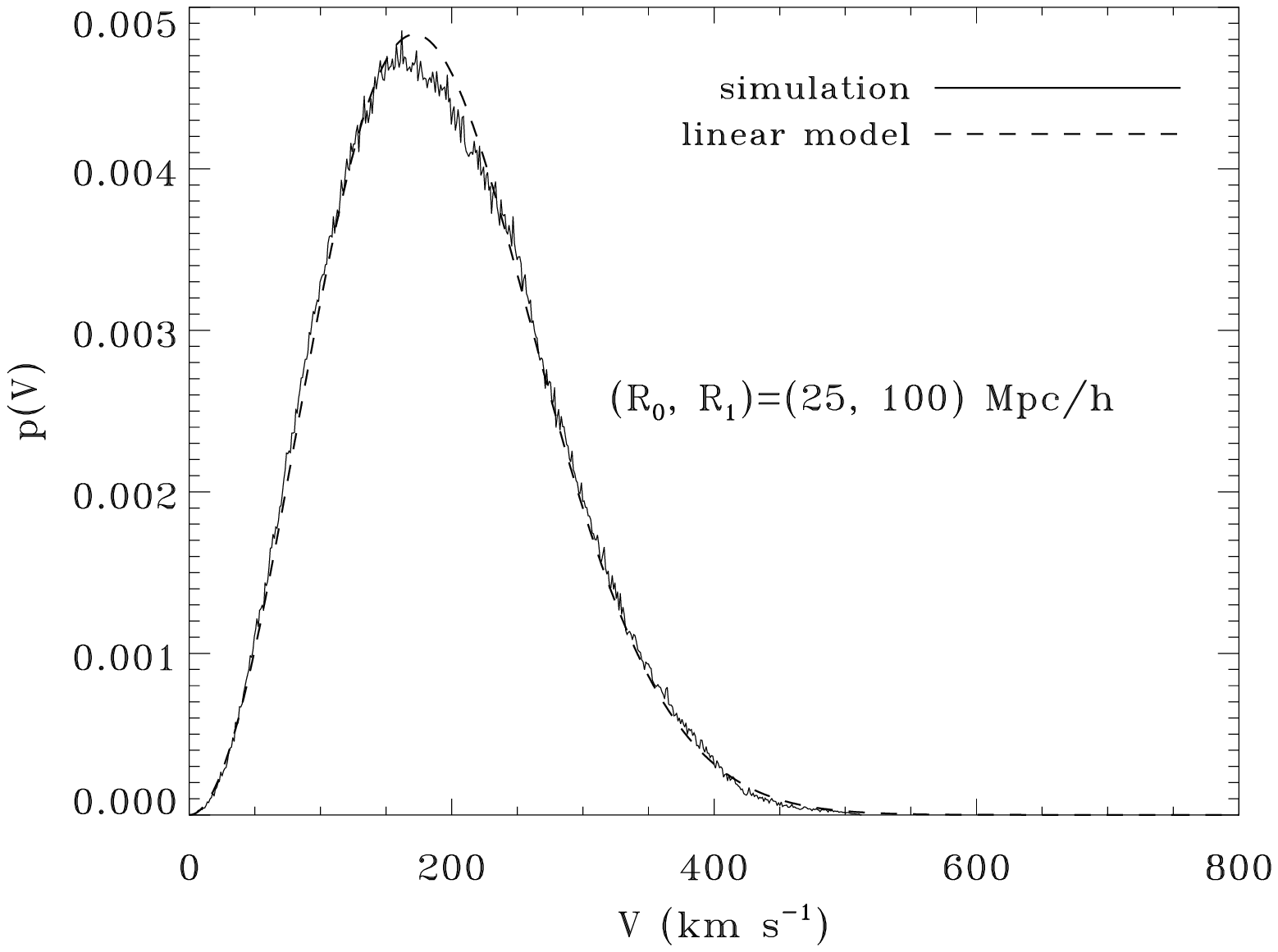}
\includegraphics{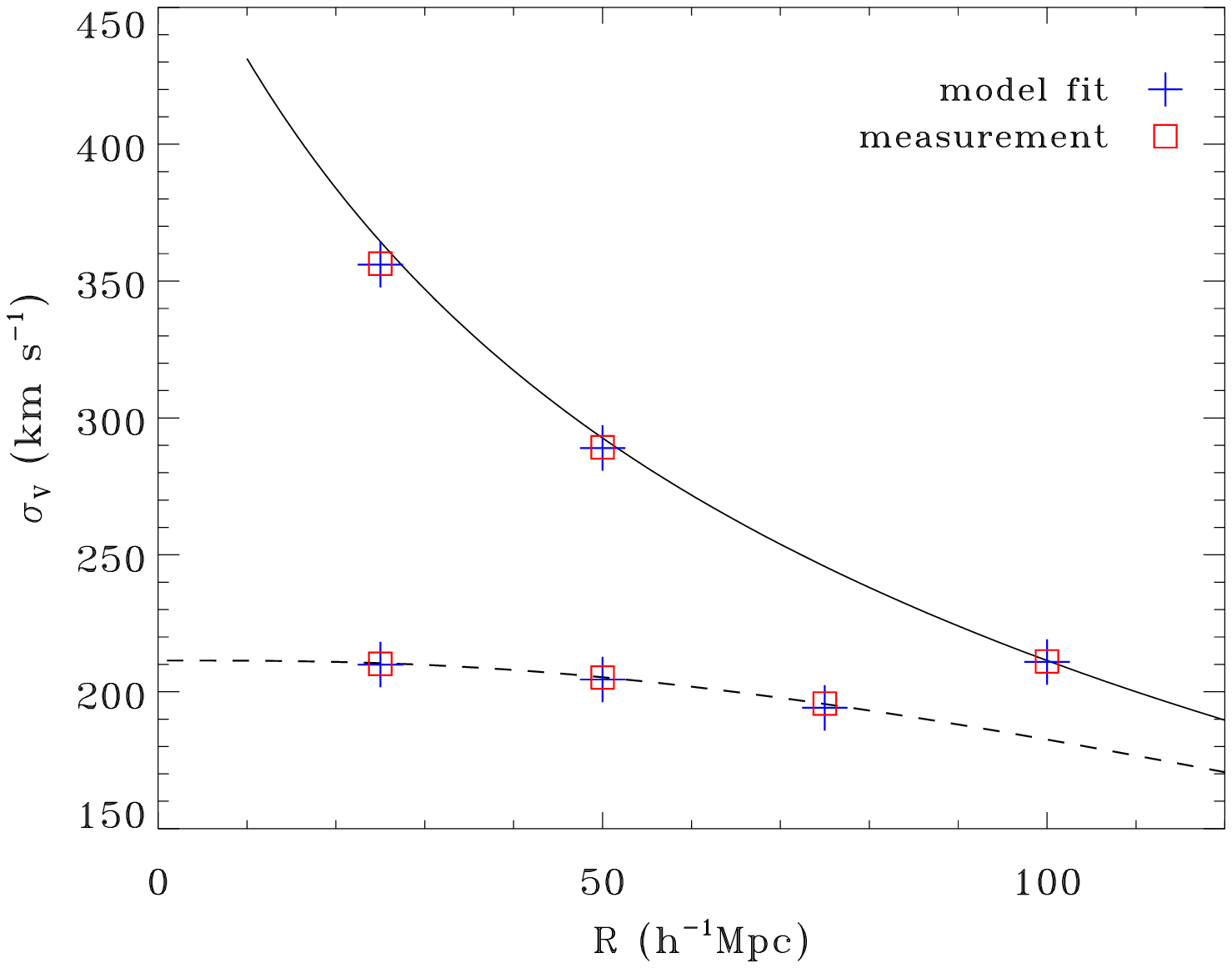}}
\caption{Left panel: $p(V)$ measured in spherical top-hat windows with different 
radius, solid lines are measurements from N-body simulation, from left to right 
with decreasing heights corresponds to 
radius of window function $R=100, 50, 25h^{-1}$Mpc respectively, dashed lines 
are prediction of the 
model of Eq.~\ref{eq:linVpdf} \&  ~\ref{eq:sigV} with 
$P_{\theta\theta}=P_{\delta\delta}^{(L)}$. 
Middle panel: $p(V)$ measured in one spherical shell window defined 
by two radius $(R_0, R_1)=(25,100)h^{-1}$Mpc.
Right panel: comparison of $\sigma_V$ predicted by the linear 
theory (solid line is of spherical top-hat window, dashed line is of 
spherical shell window with $R_0=R$, $R_1=100h^{-1}$Mpc) with estimation from 
simulation (blue crosses), and the $\sigma_V$ which provide the best fitting to 
PDFs of bulk flow in simulation with Eq.~\ref{eq:linVpdf} (red squares).
}
\label{fig:Vpdf}
\end{figure*}

To ensure fair sampling to the simulation, the over-sampling algorithm 
of \citet{Szapudi1998a} is 
implemented to generate $\sim 10^6$ cells for $R$ within $25-100h^{-1}$Mpc, 
bulk flows of halos
in random cells are estimated with Eq.~\ref{eq:vb}. In our 
measurements we adopt mainly the
spherical top-hat window, i.e. $w=1$ for all halos inside  window 
and $w=0$ otherwise, meanwhile shell window
function is also deployed for consistency check. 
Adopting the type of top-hat window function is just to simplify computation, in principle
one could try Gaussian or other more sophisticated window functions, but it will not introduce
change to the main results. Since our simulation box 
is limited in a cubic box of side length $1h^{-1}$Gpc, probe of bulk flows 
in cells of radius $>100h^{-1}$Mpc would be statistically unreliable.

Our measurements of bulk flows are displayed in Figure~\ref{fig:Vpdf}.  For 
top-hat window, $\sigma_V$
decreases with cell radius, the possibility of finding extremely large speed of 
bulk flow becomes smaller for larger volume. For shell window defined by two 
radius $R_0<R_1$, if $R_0$ is not very close to $R_1$,  
$\sigma_V(R_0, R_1)\sim \sigma_V(R_1)$. 

More importantly, PDFs of amplitude of bulk flow in simulation, no 
matter measured with spherical top-hat window or  spherical shell window, 
are all well described by the linear model. 
Agreement between simulation and model is better for larger volume as 
expected, linear theory slightly over-predicts $\sigma_V$ since 
nonlinear $P_{\theta\theta}$ is lower than $P_{\delta\delta}^{(L)}$ 
at $k \sim 0.1 h{\rm Mpc}^{-1}$ by $\sim 20\%$ 
already \citep{NusserEtal1991, CiecielgChodorowski2004, Scoccimarro2004}. 
The comparison clearly lead to the conclusion that 
to a good precision $p(V)$ obeys Maxwellian distribution which is 
completely determined by $\sigma_V$, what really matters is not the
exact shape of the widow function but rather the corresponding $\sigma_V$.
This lays out the solid ground for us to put different kinds of 
measurement together for comparison.

\subsection{The simple correction for selection function}
\label{subsec:sel}

\begin{figure*}
\resizebox{\hsize}{!}{\includegraphics{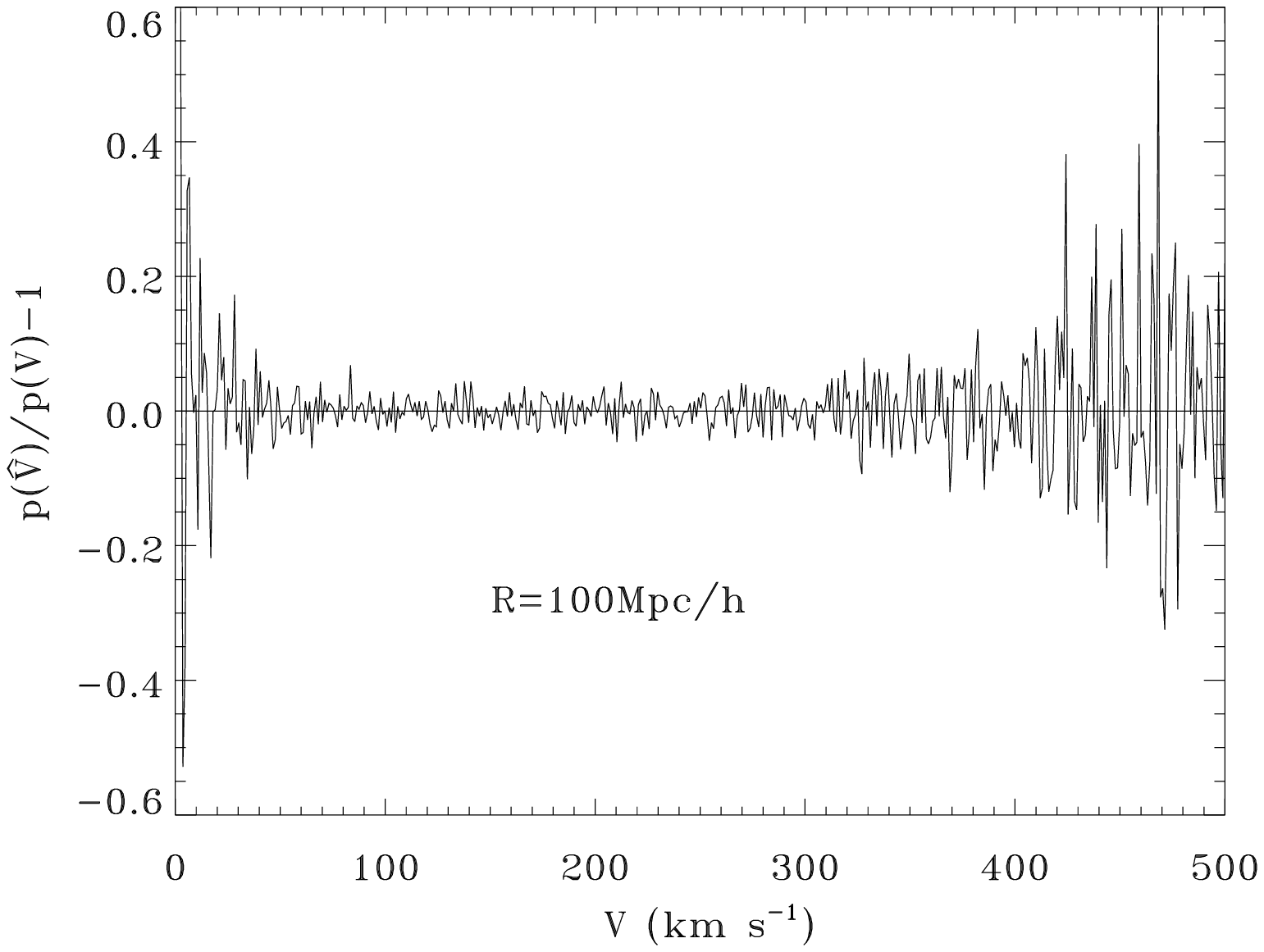}\includegraphics{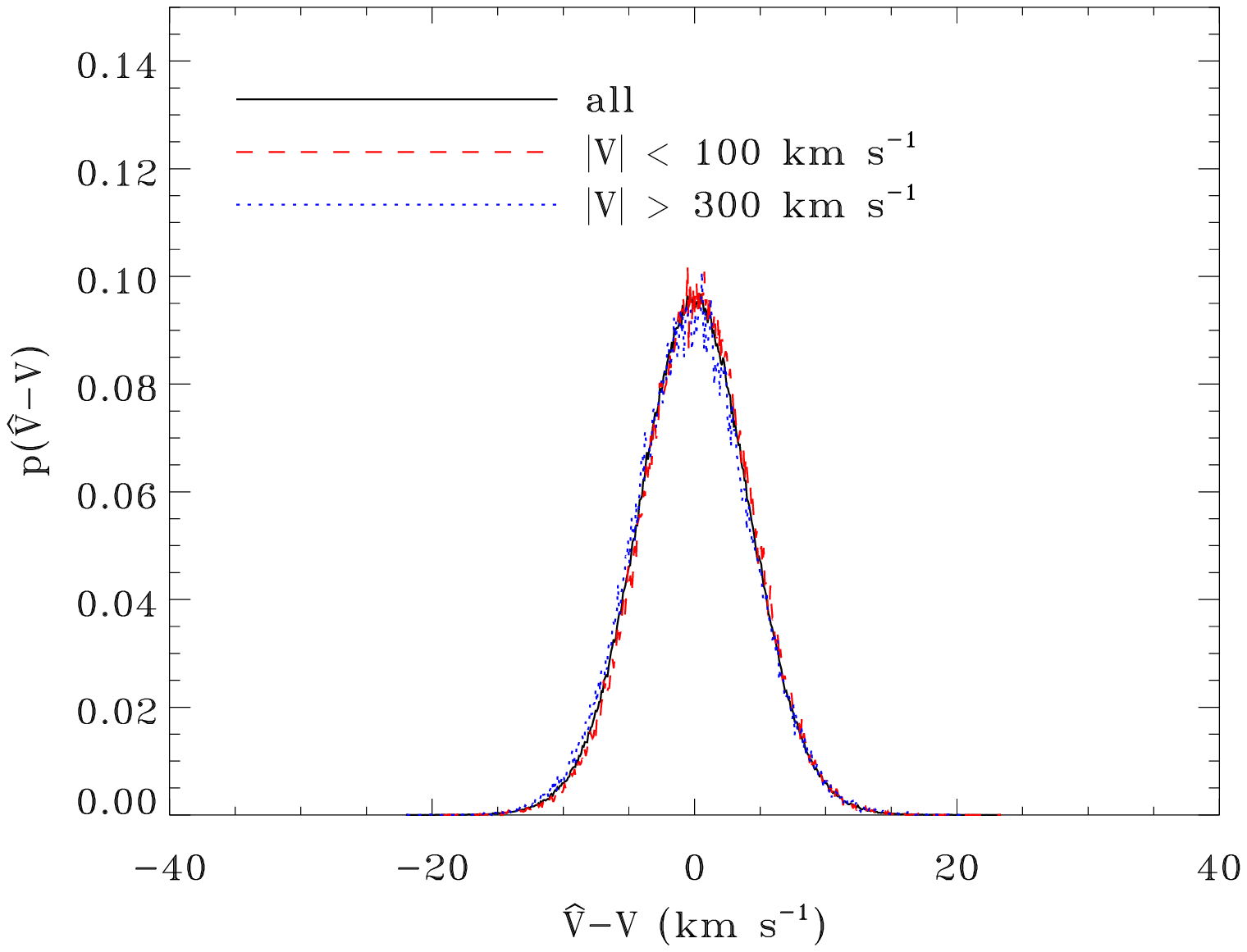}\includegraphics{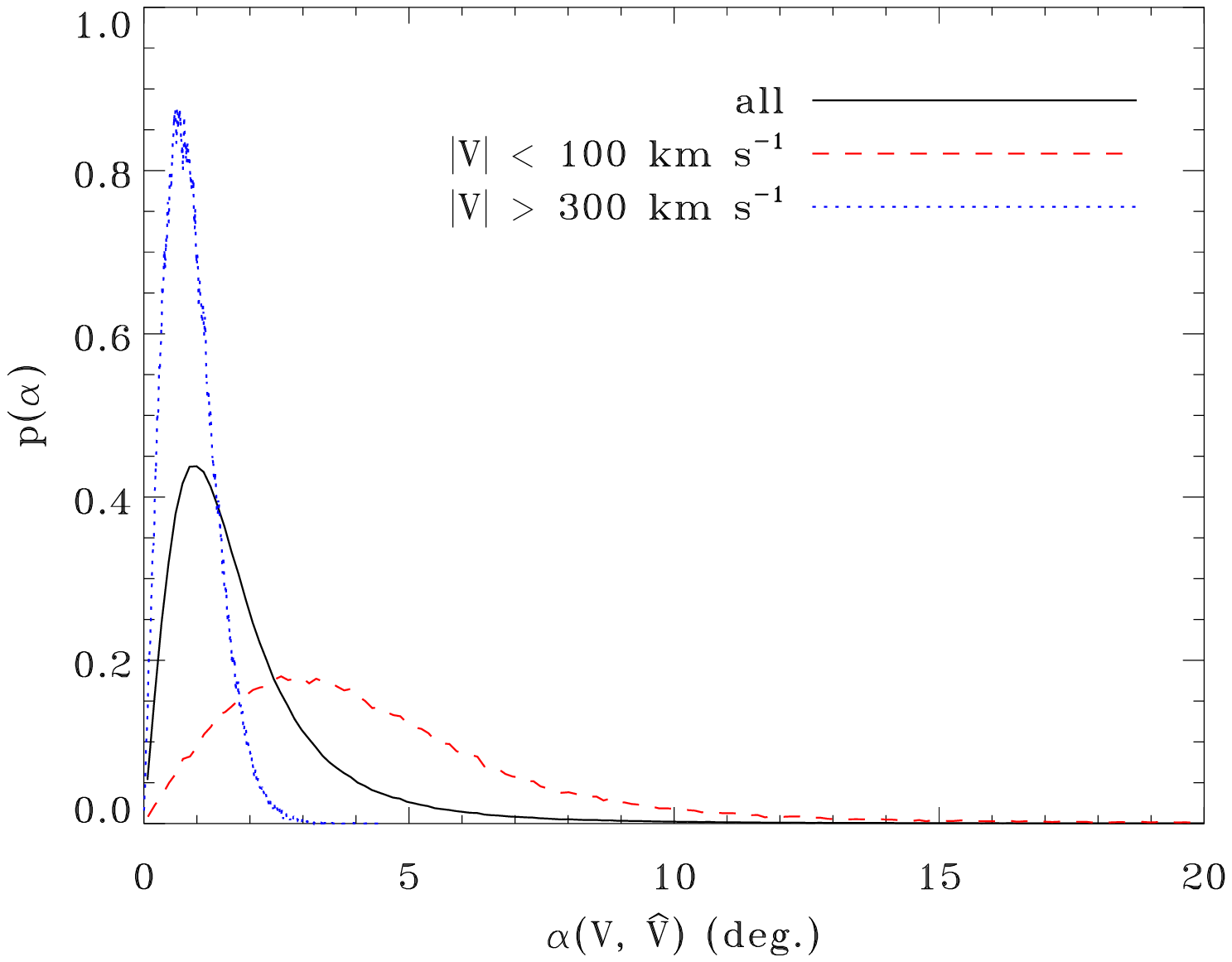}}
\caption{Effectiveness of the correction for the selection function. $\hat{\vect{V}}$ is the corrected estimation (Eq.~\ref{eq:sel})
to bulk flow with selection function applied, $\vect{V}$ is the normal one without selection function.
Left: the relative difference between $p(\hat{V})$
and $p(V)$. Middle: PDFs
of amplitude difference between $\vect{V}$ and $\hat{\vect{V}}$ in the same cell. 
Right: distribution functions of the angle between $\vect{V}$ and 
$\hat{\vect{V}}$ in the same cell, 
$\alpha=\cos^{-1} [\vect{V}\cdot \hat{\vect{V}}/(V \hat{V})]$. In the middle and the right panels 
also shown are subsamples of $V<100\kms$ (red dashed line) and $V>300\kms$ (blue dotted line).}
\label{fig:sel}
\end{figure*}

In this part we take numerical approach to assess effectiveness of the simple correction method 
of Eq.~\ref{eq:sel} for selection function. In the experiment, the sample window function is a spherical
top-hat of radius $R=100h^{-1}$Mpc, selection function is set to be in the form of the PSCz
catalogue  \citep{SaundersEtal2000},
\begin{equation}
n(r)= \left\{  
\begin{array}{lr} 
1 & \mbox{ if $r  \leq 20h^{-1}$Mpc} \\
n_* \left( \frac{r}{r_*}\right)^{1-c}\left[ 1+\left( \frac{r}{r_*}\right)^\gamma \right]^{-\beta/\gamma} & \mbox{if $r > 20h^{-1}$ Mpc}
\end{array} 
\right. \ ,
\label{eq:selsim}
\end{equation}
in which $c=1.82$, $r_*=86.4$, $\gamma=1.56$, $\beta=4.43$, and $n_*=0.397$ to give the normalization
$n(r=20h^{-1}{\rm Mpc})=1$.
During computation, for each sampling cell Monte-Carlo simulation is applied to halos in the cell
to generate its mock catalogue which radial distribution obeys with Eq.~\ref{eq:selsim}, then two 
kinds of bulk flows for each individual cell are estimated from the mock, 
one is the estimated directly with Eq.~\ref{eq:vb}, and the other is the selection function corrected 
$\hat{\vect{V}}$ by Eq.~\ref{eq:sel}. The two measurements are then compared with the results 
without selection function.

For the one given by Eq.~\ref{eq:vb}, selection function is not corrected at all, the resulting distribution of
estimated bulk speed is a Maxwellian distribution function of variance $\sigma_V=244\kms$, while the
variance of bulk flow without selection function is $211\kms$. Apparently selection function makes the sample
having reduced effective scale, inducing larger $V_p$ and $\sigma_V$. This seriously challenges the 
claim of \citet{MakEtal2011} that selection function has little influence on bulk flow estimation.

The comparison of the selection function corrected estimation with results without selection function
is displayed in Figure~\ref{fig:sel}, it appears that the simple correction of Eq.~\ref{eq:sel} can recover the
bulk flow to a good extent. The PDF $p(\hat{V})$ differs little from $p(V)$, which means that
the variance of the smoothed velocity field is actually well recovered. For individual cell, the deviation 
of $\hat{\vect{V}}$ to $\vect{V}$ is small, amplitude difference shows no systematical bias and is mainly bounded 
within $\sim 15\kms$, which seems does not vary much with the amplitude of $\vect{V}$; 
shift in direction rarely goes beyond $\sim 10^\circ$ and has most likely value of about  $1^\circ$, but the alignment
turns to be better for larger $V$.

\subsection{Bulk flow as mass weighted average of halo velocities}
\label{subsec:mv}

\begin{figure*}
\resizebox{\hsize}{!}{\includegraphics{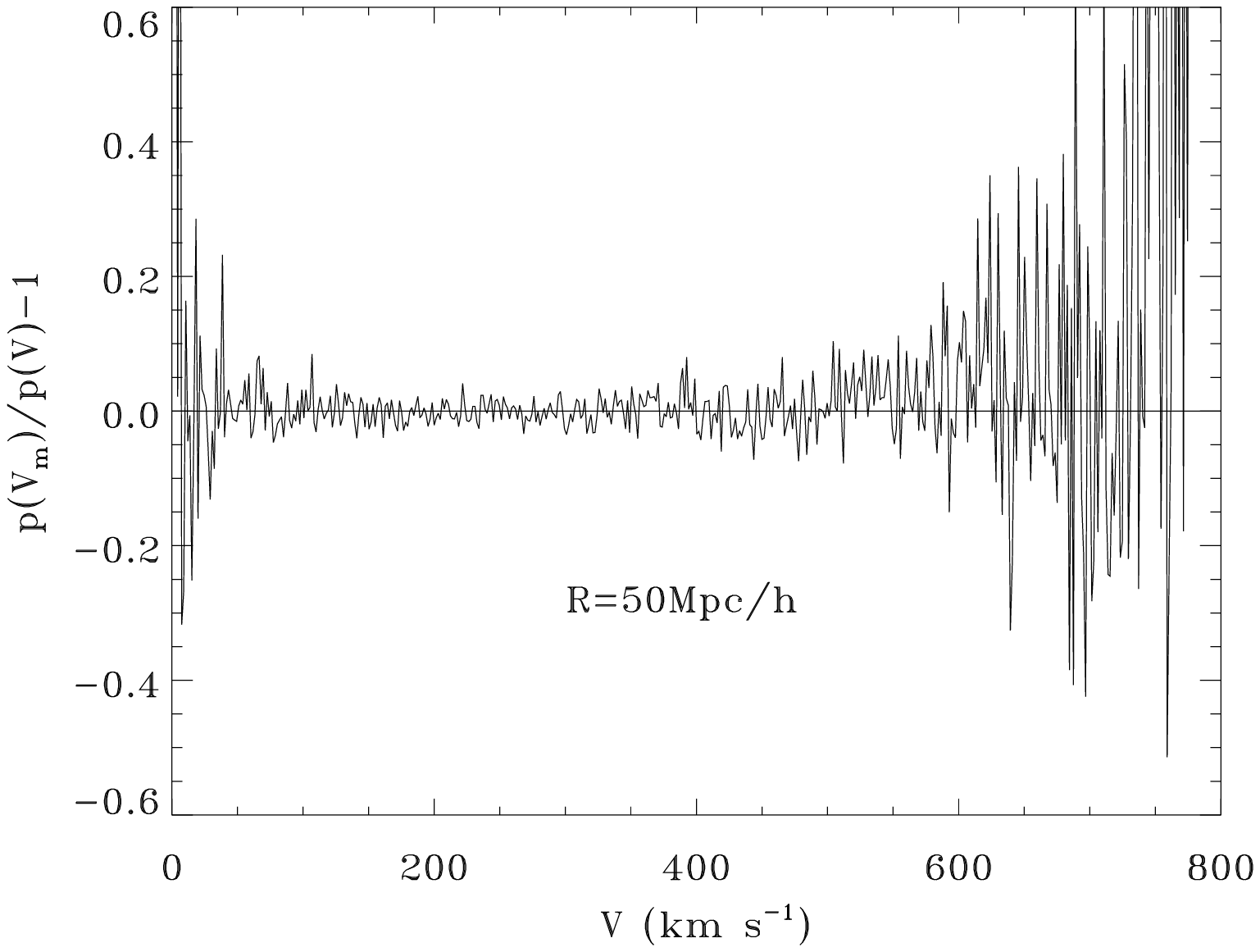}\includegraphics{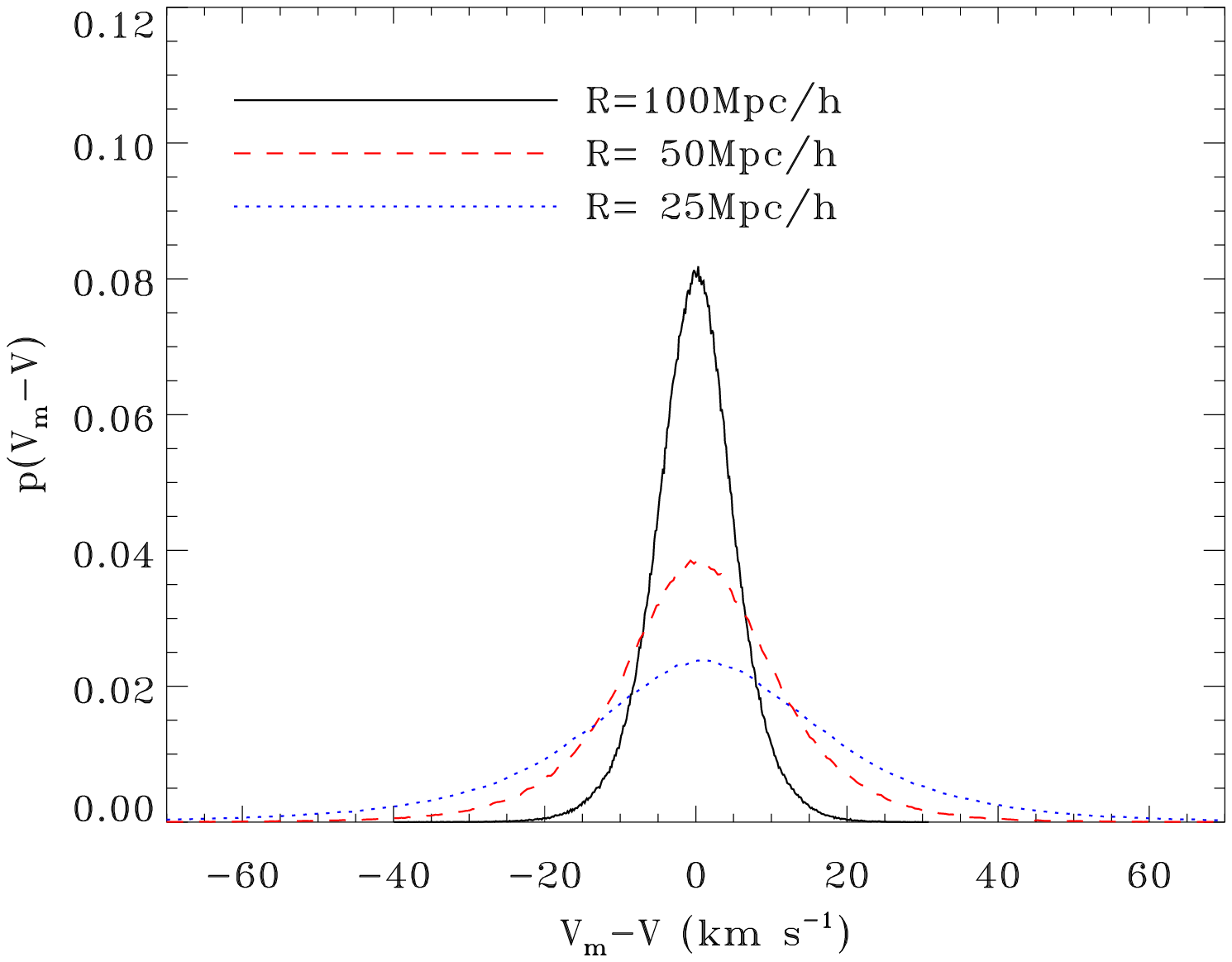}
\includegraphics{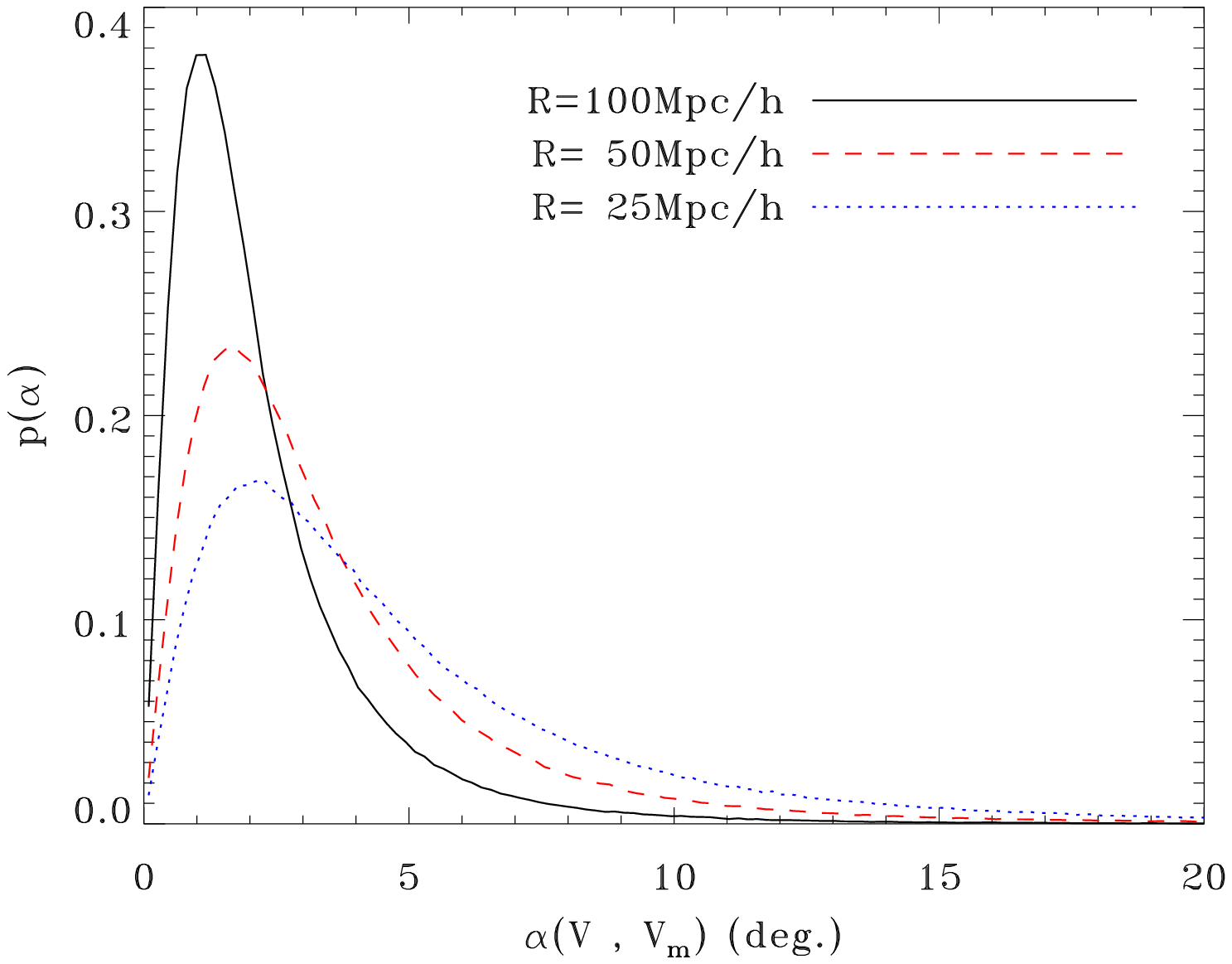}}
\caption{Differences between $\vect{V}$ and $\vect{V_m}$ in simulation. $R$ is the 
radius of the spherical top-hat window function.
Left: the relative difference between $p(V_m)$
and $p(V)$, the large fluctuation at the large $V$ is due to the almost zero 
values of PDFs at tail. Middle: PDFs
of amplitude difference between $\vect{V}$ and $\vect{V}_m$ in the same cell. 
Right: distribution functions of the angle between $\vect{V}$ and 
$\vect{V}_m$ in the same cell, 
$\alpha=\cos^{-1} [\vect{V}\cdot \vect{V}_m/(V V_m)]$.
}
\label{fig:VmV}
\end{figure*}

It is known that attenuation to CMB temperature resulted from kSZ effect
$\Delta T_{kSZ} \propto \vect{v}\cdot \hat{\vect{l}} \int n_e {\rm d} l $ in which
$\hat{\vect l}$ is the unit vector of the line-of-sight and $n_e$ is the density of free 
electrons in the galaxy cluster. 
If the aperture used to measure kSZ effect is sufficiently large, and the number of hot electrons
in the cluster can be taken for granted proportional to the mass of host halo $m_h$, 
the total kSZ effect induced temperature fluctuation will be proportional to 
$m_h \vect{v} \cdot \hat{\vect{l}}$, thus the bulk flow estimated via kSZ effect of galaxy clusters 
in fact is in principle the mass weighted average of halo velocities,
\begin{equation}
\vect{V}_m=\frac{\sum_i m_i \vect{v}_i}{\sum_i m_i}\ .
\label{eq:mv}
\end{equation}

$\vect{V}_m$ is ratio of two Gaussian random variables, the total mass 
$M=\sum_i m_i$ and momentum ${\vect P}=\sum m_i {\vect{v}_i}$, with the 
results of \citet{Pham-GiaEtal2007}
it is possible to work out a linear theoretical model for $p(V_m)$. 
Exact calculation needs knowledge of power spectra of matter, momentum and the correlation
function between matter and momentum. However a quick inspection could 
give us a rough profile.  In continuous limit $M=\sum_i m_i$
becomes $ \langle M \rangle [1+ \delta *W(R)] $,  the smoothed density contrast $\delta *W \ll 1$
if $R$ is large enough to enter the linear regime, therefore $\vect{V}_m\sim \vect{P}/\langle M \rangle$.
It has been found that
the variance of  $\vect{P} /\langle M \rangle$ is dominated by the 
$\sigma_V^2=(2\pi)^{-3}\int P_{\vect{v}\vect{v}}\widetilde{W}^2{\rm d}^3 k$ if the smoothing scale
is sufficiently large \citep{ParkPark2006},  so we can expect that $p(\vect{V}_m)\sim p(\vect{V})$. Our
results of simulation data indeed reveal that differences between $p(\vect{V}_m)$ and $p(\vect{V})$
are small (Figure~\ref{fig:VmV}). 

But in an individual cell $\vect{V}_m$ does differ from $\vect{V}$, both in 
direction and amplitude. As we can
see in Figure~\ref{fig:VmV}, $p(V_m-V)$ has width of several tens $\kms$ which decreases with larger 
cell volume. If the sample volume is small, it could appear that $V_m$ deviates from $V$
by $\sim 100\kms$ though the possibility is tiny. We also notice that $V_m-V$ does not show
apparent trend with $V$ or $V_m$. Pointing of mass weighted bulk flow 
does not coincide with $\vect{V}$. The most likely angle between them is around $1-3$ degrees for 
top-hat window of $R\in (25, 100)h^{-1}$Mpc, and becomes smaller for larger volume. Note that the
distribution of the difference angle has a rather long tail, for instance if $R=50h^{-1}$Mpc 
the probability of misalignment greater than $10^\circ$ is $\sim7.3\%$, yet not trivial. 

Hitherto only the kSZ measurements can provide estimation of mass-weighted bulk flow, meanwhile
mass weighting does not introduce significant statistical differences, which is actually supported 
by real observation \citep{LavauxEtal2012}, thus hereafter we will just
concentrate on the unweighted bulk flow.

\subsection{Halo mass dependence}
\label{subsec:sub}

\begin{figure*}
\resizebox{\hsize}{!}{\includegraphics{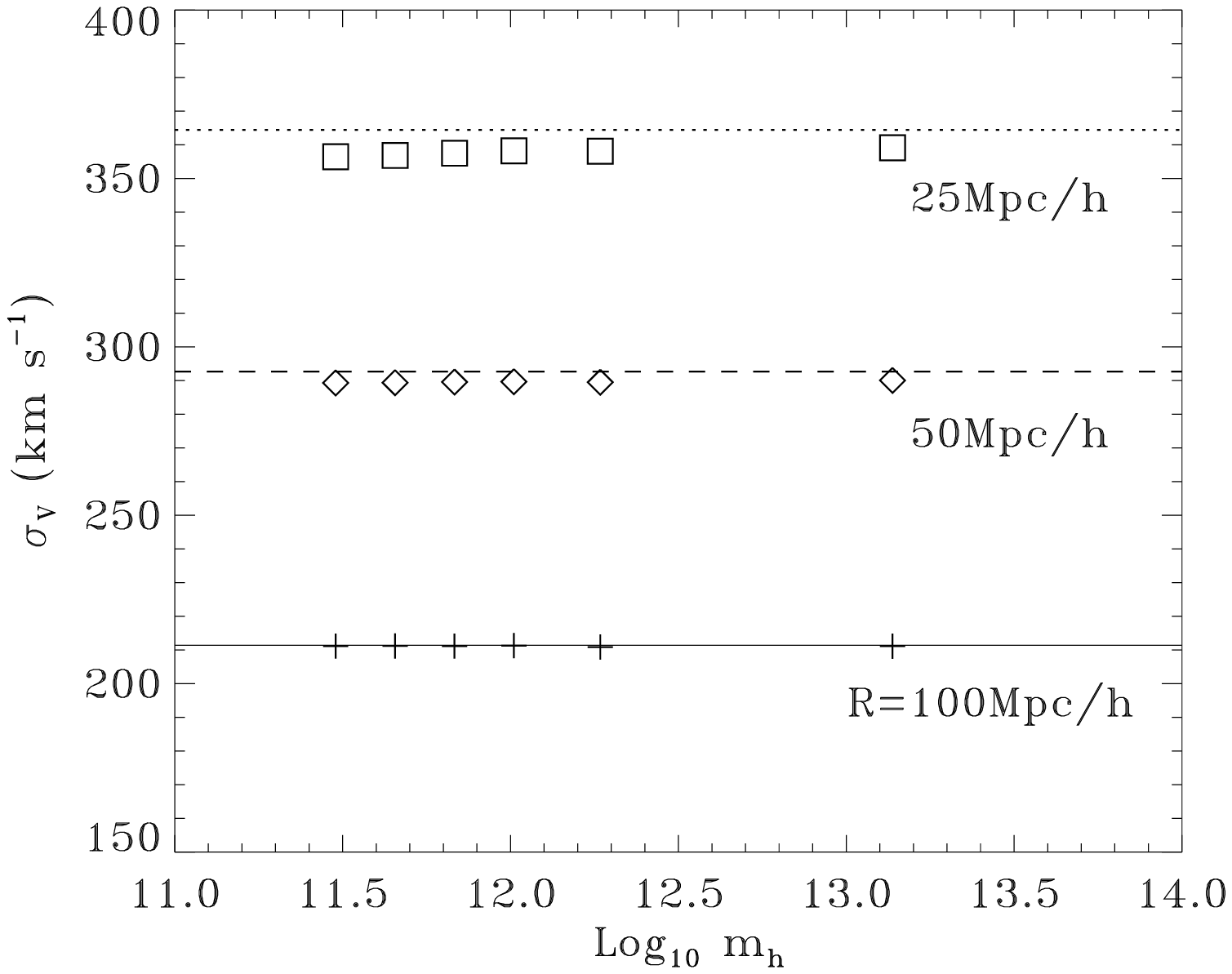}
\includegraphics{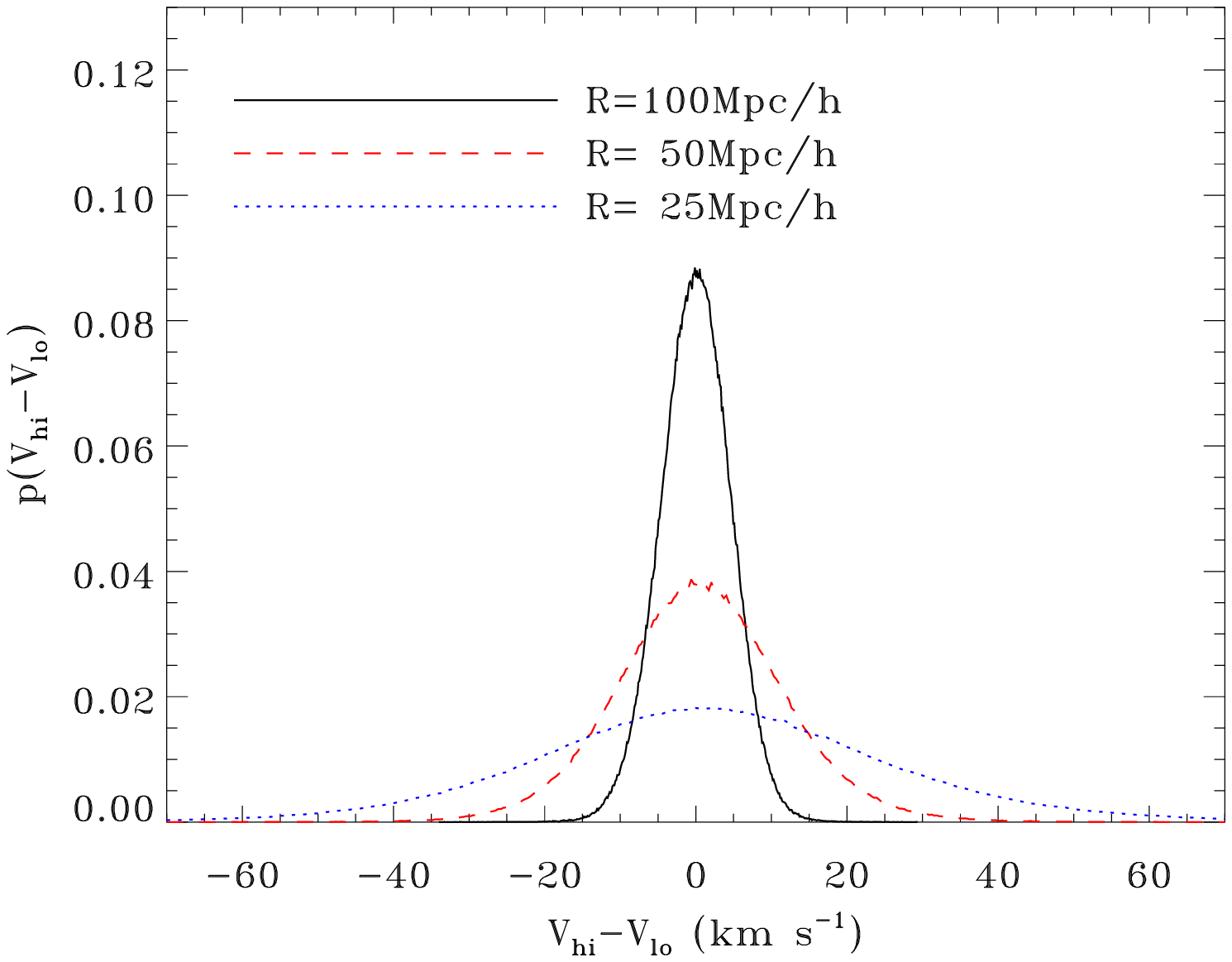}
\includegraphics{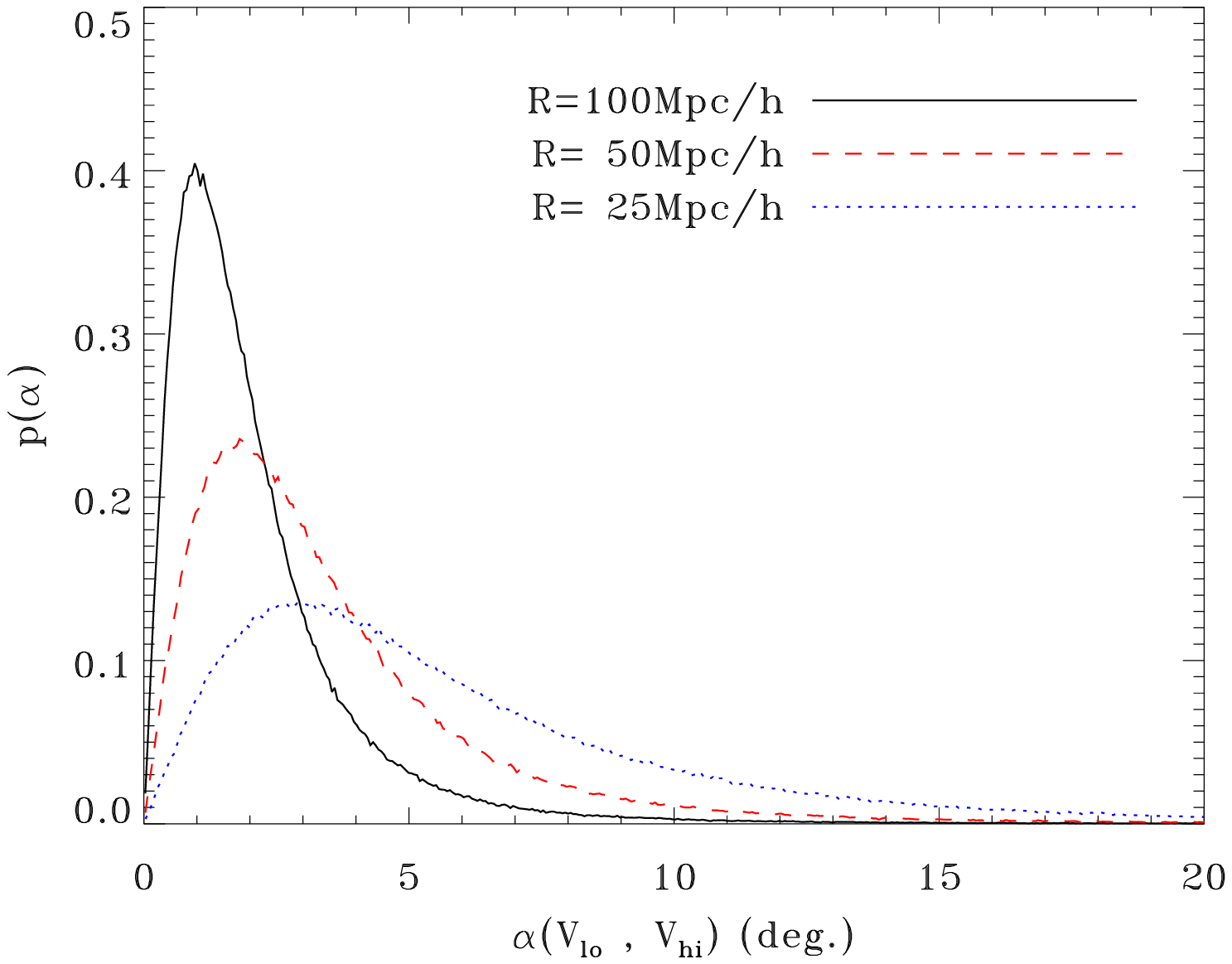}
}
\caption{Halo mass dependence of bulk flow. Left:  dependence of $\sigma_V$ on halo mass, 
symbols are measurement, their x-axis coordinates 
are the mean halo mass of subsamples, lines are
prediction of linear theory, below which are labelled with corresponding cell radius.  Middle:
distribution of amplitude difference between the bulk flow $\vect{V}_{lo}$ of the lightest halo subsample 
($\bar{m}_h=3.015\times 10^{11}\msun$)
and that of the most massive halo subsample $\vect{V}_{hi}$ ($\bar{m}_h=1.374\times 10^{13}\msun$)
in the same cell. Right: PDFs of
the angle between $\vect{V}_{lo}$ and $\vect{V}_{hi}$, $\alpha=\cos^{-1}[\vect{V_{lo}}\cdot \vect{V_{hi}}/(V_{lo}V_{hi})]$.
}
\label{fig:svmass}
\end{figure*}

There is the possibility that bulk flow may depends on the typical mass of halo sample.  The full 
halo catalogue is then divided into six subsamples by halo mass, measured $\sigma_V$ is
plotted in Figure~\ref{fig:svmass} as function of the mean halo mass of  the subsample. If the smoothing
scale is large it is obvious that there is little dependence on mass of sampled halo of bulk flow. 
For small sized windows, e.g. $R=25h^{-1}$Mpc, the measured
$\sigma_V$ of low mass subsample is slightly lower than that of high mass subsample, 
which might be just statistical fluctuation.

For individual cells,  the diversity in bulk flows measured from different halo mass bins might be non-trivial
provided that both of $R$ and $V$ are not very large (Figure~\ref{fig:svmass}). Considering the fact that
intrinsic properties of galaxies and galaxy clusters are more or less correlated with their host halo mass, 
in case that the sample depth is shallow and estimated bulk flow is of low amplitude, it would not be strange to
meet with the difficulty of achieving tight convergence among different samples.

\subsection{Consistency between observation and model}

Systematical biases in dark flow, the bulk flow measured at high redshift, are not fully
understood and precisely controlled, so we refrain ourselves from discussing high redshift case. 
Most of the local (or nearby) bulk flow measurements has redshift less than $\sim 0.06$, resulting redshift evolution
of $\sigma_V$ with respect to $z=0$ is of magnitude of a few percents at most, which can be
comfortably ignored. Window functions in different works are not the same at all, but the excellent
performance of the linear model provides an unified scheme. Since PDFs of bulk flow is solely determined
by $\sigma_V$, independent of the type of the window, the radius of a top-hat window which gives
the same linear $\sigma_V$ as the window function used in observation can acts as the 
effective scale corresponding to a particular sample. 

To check if an observed bulk flow is consistent with $\Lambda$CDM model, we need to figure 
out the variance ranges of $V$. The most likely amplitude of $V$ is $V_p=\sqrt{2/3}\sigma_V$,
derived via ${\rm d}p(V)/{\rm d}V=0$, and the variance range at different levels are computed 
through Eq.~\ref{eq:range}. 
Given significance levels of $1\sigma$, $2\sigma$ and $3\sigma$, corresponding confidence
probabilities are $\epsilon_{1,2,3}=68.3\%, 95.5\%, 99.7\%$ respectively, we choose to define
variance range $\Delta V$ of $V$ around $V_p$ at specified level through
\begin{equation}
P(|V-V_p|\geq \Delta V_{1,2,3})\leq 1-\epsilon_{1,2,3}\ .
\label{eq:var}
\end{equation}
Since by definition $V\equiv |\vect{V}| \geq 0$ and the probability 
$\int_0^{2V_p} p(V){\rm d}V = 95.4\%<\epsilon_{2,3}$, 
to stick with Eq.~\ref{eq:var} variance ranges at $2\sigma$ and $3\sigma$ levels shall be 
translated to $(0,V_p+\Delta V_{2,3})$ in that
$\int_0^{V_p+\Delta V_{2,3}}p(V){\rm d}V =\epsilon_{2,3}$, while the $1\sigma$ variance range is
the usual one,  $(V_p-\Delta V_1, V_p+\Delta V_1)$ with 
$\int_{V_p-\Delta V_1}^{V_p+\Delta V_1} p(V) dV=\epsilon_1$.
Numerical computation with Eq.~\ref{eq:range} then tells that $\Delta V_1=0.38875\sigma_V$, 
$\Delta V_2=0.81904\sigma_V$ and $\Delta V_3=1.35577\sigma_V$.

\begin{figure}
\resizebox{\hsize}{!}{\includegraphics{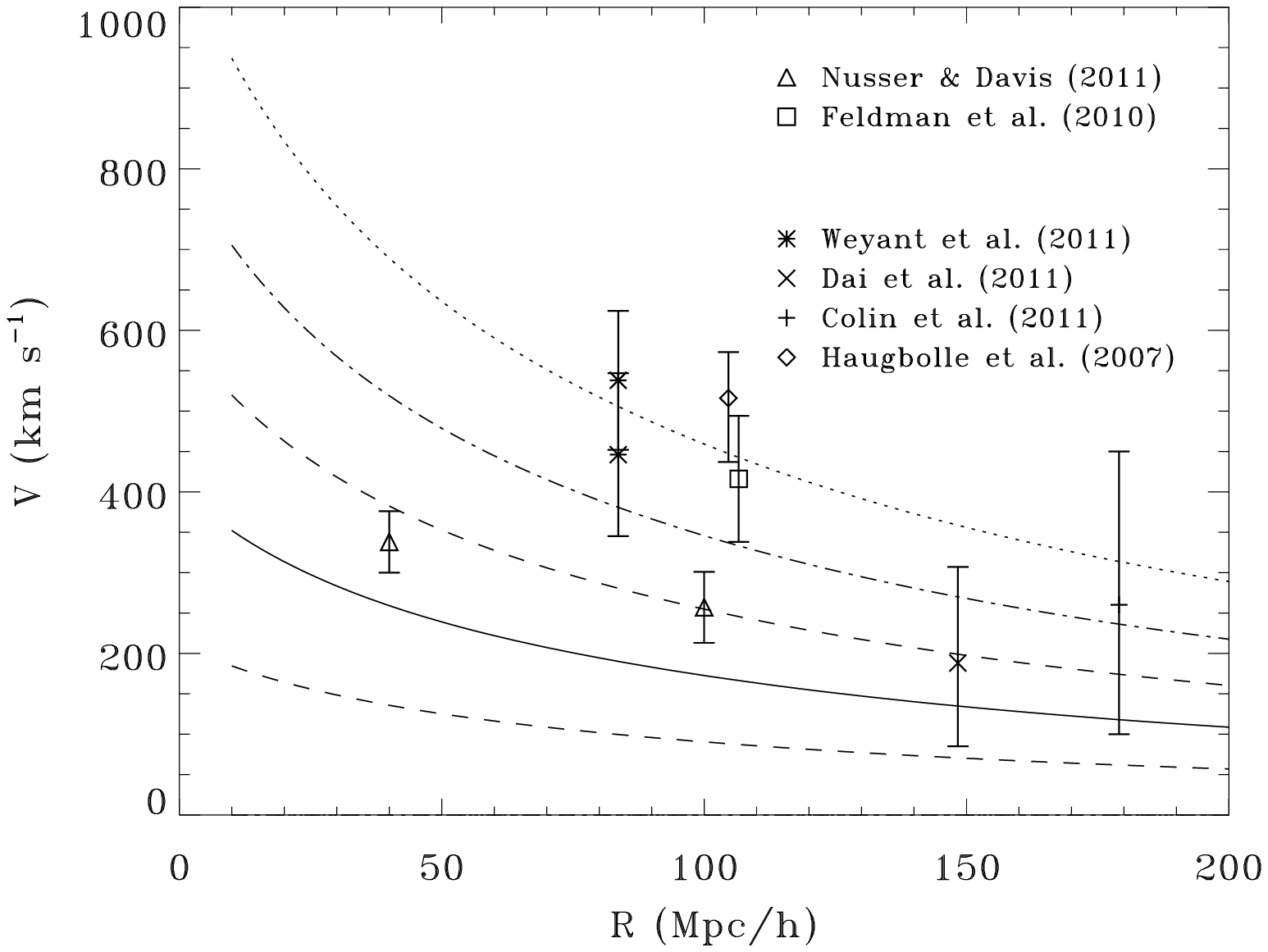}}
\caption{Measured local bulk flows against prediction of linear model. Symbols are some recently estimated
amplitudes of local bulk flows from observational data of galaxies \citep{FeldmanEtal2010, NusserDavis2011}
and supernovae \citep{HaugbolleEtal2007, ColinEtal2011, DaiEtal2011, WeyantEtal2011}. Solid line is
the model predicted most likely bulk speed, dashed lines indicate variance at $1\sigma$ level ($68.3\%$), 
dot-dashed line and dotted line are at levels of $2\sigma$ ($95.5\%$) and $3\sigma$ ($99.7\%$) respectively.}
\label{fig:obs}
\end{figure}

Several recent measurements from observational data are over-plotted upon model prediction
in Figure~\ref{fig:obs}. One has to keep in mind that our calculation of effective scales
has assumed that all other factors affecting estimation have been perfectly corrected, such as 
non-linearity, selection function and sky coverage incompleteness. 
Construction details of these samples are
often too sketchy to render appropriate weights for our calculation. 
However the imperfection actually reduces
their effective volumes so that the true effective scales will be
smaller, i.e. the data points will shift leftward along horizontal axis 
in Figure~\ref{fig:obs}. So we shall
deem Figure~\ref{fig:obs} as the mostly conservative judgment of the consistency between 
observation and theory.
Nonetheless, from Figure~\ref{fig:obs}, it appears that observation 
of our local Universe does not rule out the $\Lambda$CDM model. Those results
of \citet{HaugbolleEtal2007},  \citet{FeldmanEtal2010} and \citet{WeyantEtal2011} that often quoted
as supporting evidence disfavoring standard $\Lambda$CDM model are
around $3\sigma$ level, but the significance will be smaller if error bars are
taken into account. In addition, considering that many other measurements 
\citep[including the not officially published report of][]{Wang2007} are in 
fact consistent with $\Lambda$CDM model, we prefer to choose conservative
standpoint on the issue.

\subsection{Bulk flow and mass distribution in the cell}

It is interesting to investigate the relation between bulk flow and the 
mass distribution
in the sample volume, one might wonder whether one could infer bulk flow from
mass distribution if peculiar velocity data is absent, since in linear regime 
Fourier modes
of velocity field can be derived from modes of the density field. However 
from Eq.~\ref{eq:vdef}
it is clear that bulk flow is determined by those modes of wavelengths larger than the
characteristic scale of the window function, if the Fourier transformation of 
the density field is restricted to the same volume in which bulk flow is measured, those modes 
of long wavelength accounted for bulk flow are missing. In fact it has been clearly shown
by \citet{NusserDavis1994} that bulk flow is completely immune to internal mass distribution.

Our measurements confirm the expectation. The first quantity we checked is the 
mass monopole, the total mass ($\sum_i m_i$) or the total number ($N$) of halos 
in the volume, which is equivalent to the density fluctuation smoothed
by the window function. 
Correlation coefficients are computed 
to denote the correlation strength between amplitudes of bulk flow and mass 
monopole (Table~\ref{tab:corr}). Apparently bulk flow is not correlated with
the total mass and the total number of halos in the volume at all.

\begin{figure}
\resizebox{\hsize}{!}{\includegraphics{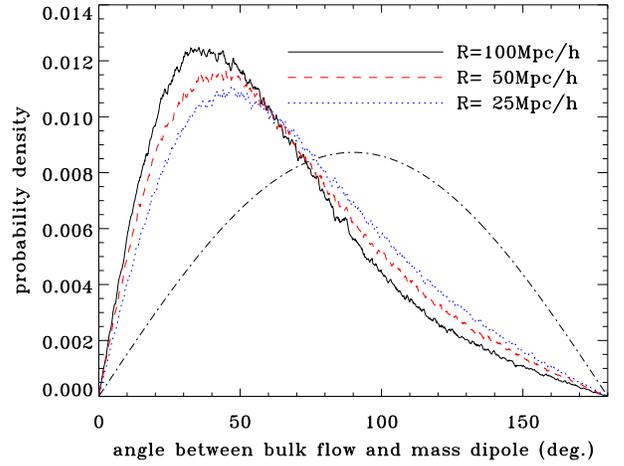}}
\caption{Probability distribution function of the angle between bulk flow and 
mass dipole. Dash-dot line is the expectation of null correlation.}
\label{fig:velmass_ang}
\end{figure}

Two kinds of dipoles of halo distribution are measured, 
$\langle \vect{r}_i/r_i \rangle $ and the mass weighted
one $(\sum m_i\vect{r}_i/r_i)/\sum m_i$. Our results show that including halo 
mass or not makes little
difference. As can be seen in Table~\ref{tab:corr} and Figure~\ref{fig:velmass_ang}, 
mass dipole correlates with bulk flow very weakly both in amplitude and 
direction. The correlation becomes 
slightly tighter as sample volume increases, the peak of the distribution 
function of the misalignment angle between mass dipole and bulk velocity is 
at $\sim45^\circ$ for $R=25h^{-1}$Mpc while shifts
to $\sim 32^\circ$ for $R=100h^{-1}$Mpc (Figure~\ref{fig:velmass_ang}).

\begin{deluxetable}{crrrrrr}
\tablecolumns{7}
\tablecaption{Correlation coefficients between amplitude of bulk flow and mass distribution}
\tablewidth{0pc}
\tablehead{
\colhead{Cell radius} & \colhead{} &
\multicolumn{2}{c}{Mass monopole}  & \colhead{} & \multicolumn{2}{c}{Mass dipole}\\
\cline{1-1} \cline{3-4}  \cline{6-7} \\
\colhead{R ($h^{-1}$Mpc}) & 
\colhead{} &
\colhead{$\sum_i m_i$} & 
\colhead{$N$ } &
\colhead{} &
\colhead{$\frac{\sum m_i\vect{r}_i/r_i}{\sum m_i}$} &
\colhead{$\langle \frac{\vect{r}_i}{r_i} \rangle$} }
\startdata
25 & & 0.024 & -0.0004 & $\ldots$ &  0.139  & 0.134   \\
50 & & 0.004 & -0.0188 & $\ldots$ & 0.188  & 0.187   \\
100 & & 0.014 & -0.0031 & $\ldots$ & 0.245  & 0.249   \\
\enddata
\label{tab:corr}
\end{deluxetable}

\section{The fastest bulk flow}

There are some works claim detection of unusually large 
bulk flow \citep[e.g.][]{FeldmanEtal2010, WeyantEtal2011}, it is interesting to 
check properties of these special parts in
$\Lambda$CDM universe. 
The largest amplitudes of bulk motion measured in our simulation 
for $R=25, 50, 100h^{-1}$Mpc 
are $1070, 778, 514\kms$ respectively, the one for $R=100h^{-1}$Mpc 
is already very close to those observational results. 
The possibility of residing in the cell having the fastest bulk motion is
defined by the ratio of the cell volume to
the total volume of simulation, $\approx 0.42\%$ for $R=100h^{-1}$Mpc. 
However, given the diameter of the cell as large as $200h^{-1}$Mpc against the 
simulation box length $1h^{-1}$Gpc, 
a huge volume moving at speed more than $\sim 500\kms$ will yield
observable features too prominent to be missed.

\begin{figure}
\resizebox{\hsize}{!}{\includegraphics{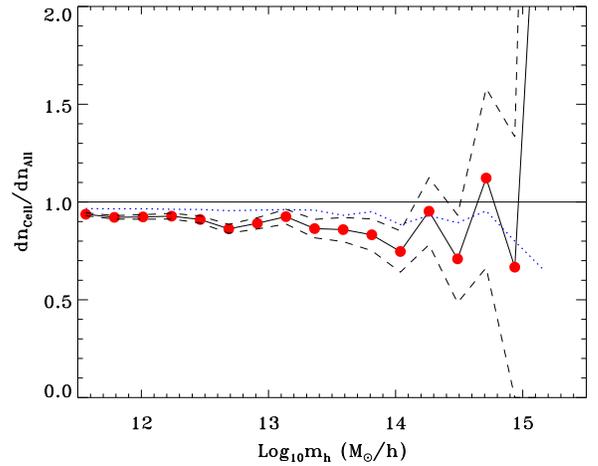}}
\caption{Difference in halo abundance between the fastest cell 
with $R=100h^{-1}$Mpc and the full
halo catalogue, ${\rm d}n_{Cell}/{\rm d}n_{All}$, ${\rm d}n_{Cell}$ 
is the mass function of halos in
the fastest cell, ${\rm d}n_{All}$ is of the full halo catalogue.
Red solid circles connected by solid line is of the fast cell, which is enclosed
by dashed lines marking its Poisson variance. The blue dotted line is the average
of the top-ten fastest cells. The horizontal solid line is 
of ${\rm d}n_{Cell}/{\rm d}n_{All}=1$.}
\label{fig:massf}
\end{figure}

In this report we will choose the $R=100h^{-1}$Mpc case as example to study
peculiarity of the cell showing fastest bulk motion.
The first physical quantity checked is the halo mass function in the cell, 
for comparison halo mass functions of the top-ten fastest cells 
(centers separated by at least $100h^{-1}$Mpc) are also measured.
Shown in Figure~\ref{fig:massf} are measured cell halo mass functions divided by 
the halo abundance in the full catalogue. 
Among the ten cell mass functions, most (more than 7 of 10) are smaller than
the halo mass function of the full catalogue for $m_h < \sim 10^{14} \msun$)
In high mass regime due to the very small number of 
high mass halos we can not withdraw any reliable conclusions though the mass 
function of the fastest cells demonstrates a high tail.
So far we would only cautiously conclude that in high bulk flow regions there is 
the tendency of finding less number of small mass halos.

\begin{figure*}
\resizebox{\hsize}{!}{\includegraphics{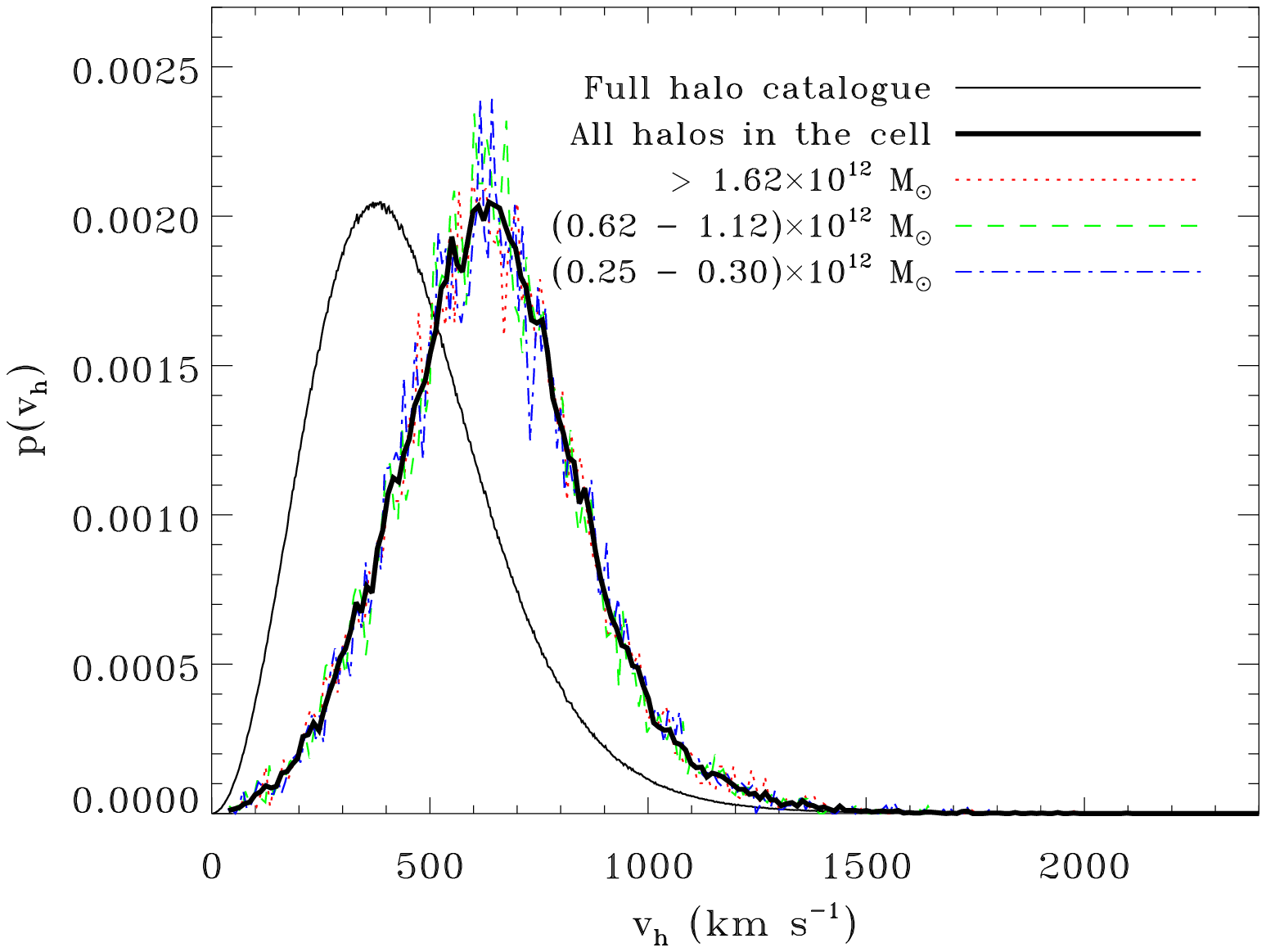}\includegraphics{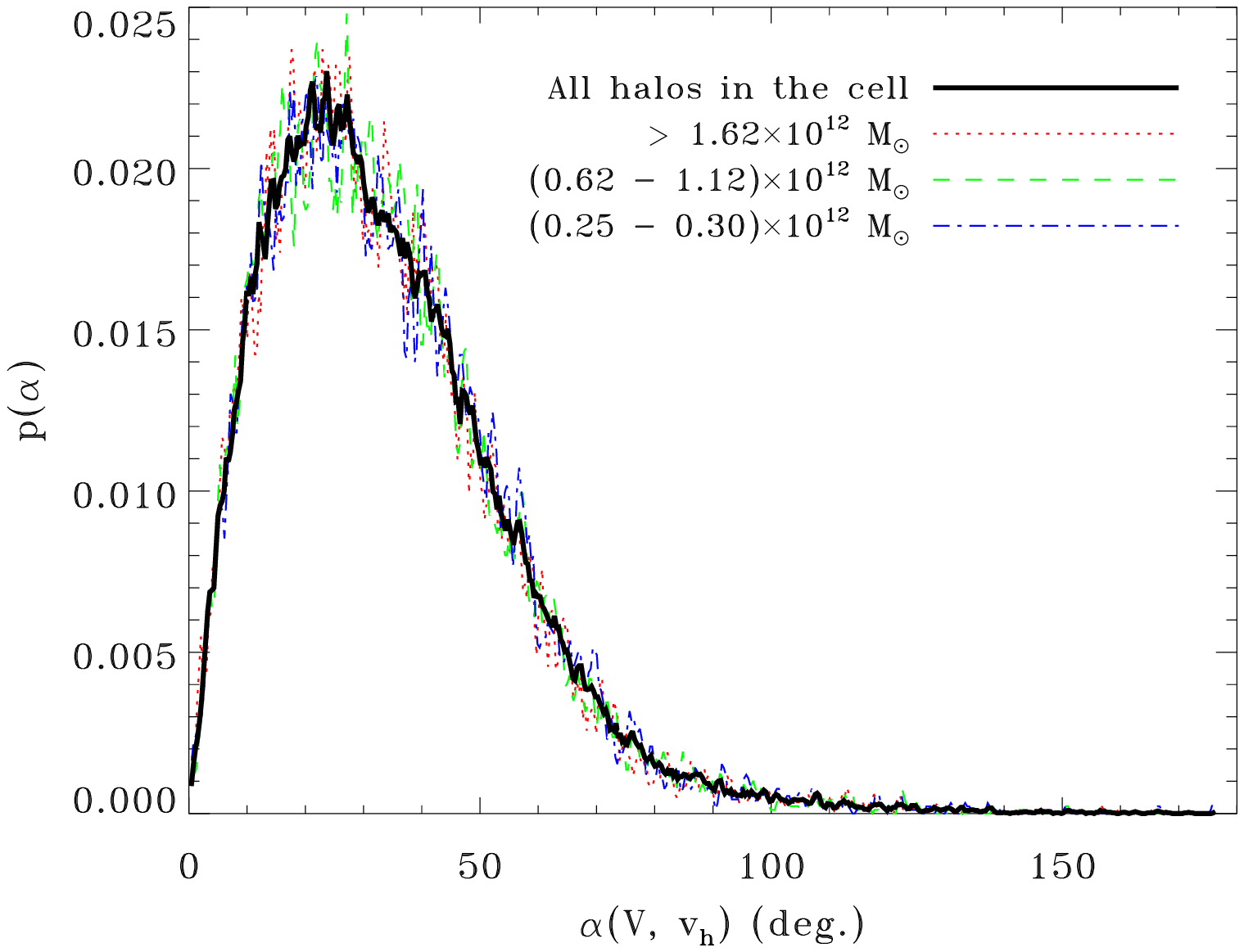}}
\caption{Velocity distribution of halos in the cell with the largest 
bulk flow ($R=100h^{-1}Mpc$). 
Left panel is the distribution of amplitudes of halo velocities, think solid line is
of halos in the cell, think solid line is of the full halo catalogue, other color
lines are of halos in three different mass bins as labelled in figure legend.
Right panel presents distributions of the angle between bulk flow $\vect{V}$ and halo 
velocity $\vect{v}_h$, $\alpha=\cos^{-1} [\vect{V}\cdot \vect{v}_h/(V v_h)]$.}
\label{fig:vaincell}
\end{figure*}

Distribution of all halo velocities in the cell is illustrated in 
Figure~\ref{fig:vaincell}, also shown are
velocity distribution functions of these halos in mass bins
$(0.25-0.3)\times10^{12}\msun$, $(0.62-1.12)\times10^{12}\msun$ and $>1.62\times10^{12}\msun$. 
Mass binned halos do not exhibit any significant differences in aspect of velocity 
distribution, which eases the worry of possible bias in mass selected halo samples.
Distribution function of the angle between halo velocity and bulk flow is 
very skewed toward small misalignment, the peak is around $20\sim30^\circ$ 
but not the $0^\circ$, about $90\%$ halos are moving in direction within $60^\circ$ to 
the bulk flow. It appears that halos in the cell with largest bulk velocity are 
more likely to have higher speed, the peak of velocity amplitude distribution of 
halos in the cell is at around $600\kms$
while that of the full halo catalogue is at $\sim 380\kms$ (Figure~\ref{fig:vaincell}).

\begin{figure}
\resizebox{\hsize}{!}{\includegraphics[angle=90]{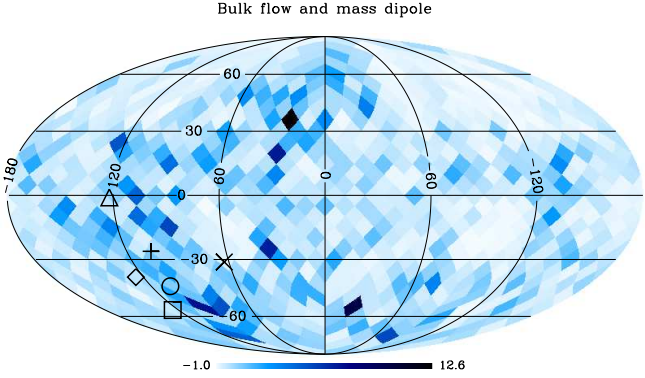}}
\caption{Mollweide projection of directions of the largest bulk flow and mass dipoles. 
Mass dipoles are computed from the mass field represented by halos, $\sum m_i \vect{r}_i/r_i$.
The plus symbol marks the direction of the largest bulk 
flow ($514\kms$, cell radius $R=100h^{-1}$Mpc), 
the cross symbol is the direction of dipole of halos in the cell, the angle
between them is $37.4^\circ$. 
Triangle, square, diamond, and circle
symbols are directions of mass dipoles within shells
$(100, 150), (150, 200), (200, 250), (250,300)h^{-1}$Mpc, 
which deviate from the bulk flow by about
$30^\circ$, $33^\circ$, $21^\circ$, $17^\circ$ respectively.
The color map is the surface mass density contrast of halos
in the cell, plotted in the Healpix scheme (dark color for high density contrast). }
\label{fig:massd}
\end{figure}

It has been examined that bulk flow basically is weakly correlated with the 
internal mass dipole on average. But for the cell with the largest bulk flow, 
intuitively one would conjecture there should be certain very massive 
clumps neighboring to the cell, their gravitational action may play a dominant 
roll in causing such extreme bulk flow of nearby halos. As an attempt to justify 
the paradigm, mass dipoles in shells within $R=100-300h^{-1}$Mpc to center of 
the $100h^{-1}$Mpc cell with largest bulk flow are calculated in four layers
with the help of the Healpix package \citep{GorskiEtal2005}, 
if there is unusual distribution of matter in a layer, mass dipole of the 
layer will be the efficient indicator. Projected directions of bulk flow and mass 
dipoles are displayed in Figure~\ref{fig:massd}. When shell moves outward
mass dipole pointing walks fairly randomly around the bulk flow, the misalignment 
angle varies between $\sim 20-40^\circ$ which is analogous to the typical value in
Figure~\ref{fig:velmass_ang}. It seems that dipoles of local environmental mass 
are only aligned crudely with the bulk flow. The correlation is not negligible, however
since these misalignment angles are not small, we have no strong support 
from the simulation to attribute the 
extremely large bulk flow in such a huge volume mainly to inhomogeneous environment.

\begin{figure}
\resizebox{\hsize}{!}{\includegraphics{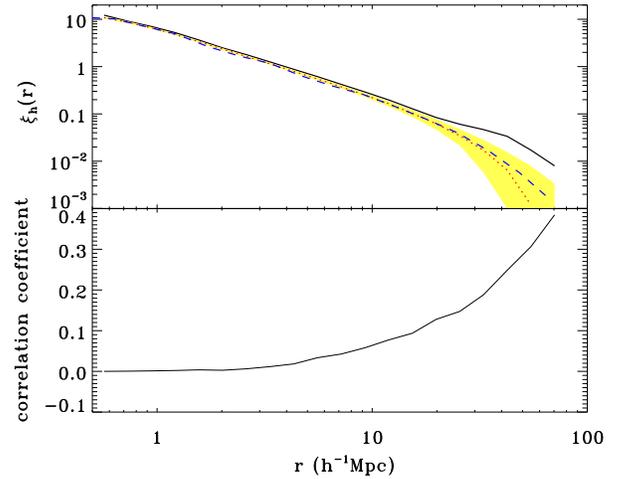}}
\caption{Correlation between bulk velocity and clustering of halos in the cell,
cell radius is $R=100h^{-1}$Mpc.
Top panel: solid line is the two-point correlation function $\xi_h$ of halos
in the cell with the largest bulk velocity, dotted line in red
is the average over 1000 $\xi_h$s measured in randomly selected cells
while the yellow shadow marks the corresponding 1$\sigma$ variance, dashed line
in blue is the result of all halos in our halo catalogue. Bottom: correlation
coefficients between amplitude of bulk velocity and $\xi_h(r)$ as function of $r$,
summarized from measurements of 1000 randomly selected cells.}
\label{fig:xi}
\end{figure}

Anomaly in clustering of halos in the special cell is detected (Figure~\ref{fig:xi}),
halo two-point correlation functions of 1000 
randomly located cells of radius $100h^{-1}$Mpc and the full halo catalogue 
are also calculated. $\xi_h$ averaged over the 1000 measurements agrees with $\xi_h$ of the 
full halo catalogue at scales $r\lesssim 30h^{-1}$Mpc, then drops down more quickly to zero
at larger scales due to the integral constraint resulted from finite volume of 
cell \citep{LandySzalay1993}. 
As we are interested in $\xi_h$ in a finite volume, we did not bother ourselves
to apply relevant correction. $\xi_h$ of the cell with largest bulk velocity is higher
than the average of random cells by around $2\sigma$ at scales $r \lesssim 20h^{-1}$Mpc, 
at larger scales the excess of clustering power rises to level of $\sim2-3\sigma$. 
In order to assess the statistical significance of the event, correlation
coefficients of bulk flow amplitude and $\xi_h(r)$ are computed from the
1000 random cells (bottom panel in Figure~\ref{fig:xi}). Little correlation is
detected at scales $r\lesssim 20h^{-1}$Mpc, then 
at larger scales the correlation becomes much stronger, indicating that
in cells with high bulk velocity it is truly more possible to find power excess in halo
clustering at large scales, which appears to be in line with the findings
in \citet{MacaulayEtal2011}. We further check the halo two-point correlation functions
in the top-ten fastest cell, and we find that 7 of the 10 demonstrate power excess at similar scales.
Note that the power excess at large scales can not be 
ascribed to integral constraint, for the leading term of integral constraint in this regime 
is negative and will bend $\xi_h$ downward \citep{LandySzalay1993}. 
 
\section{summary and discussion}
Through analysis of the Pangu simulation,
it is confirmed that bulk flow of halos follows Maxwellian distribution which 
is completely determined by a single parameter,the bulk velocity dispersion. 
We find that the dispersion measured in simulation agrees with the prediction 
of linear perturbation theory of structure formation very well, non-linearity
only becomes important when the sampling volume is very small. 
In most cases mass weighted bulk flow has some minor
statistically differences to the unweighted one, but the will not affect the
overall statistics significantly. It is also revealed that statistically bulk flow
has little systematical dependence on the mass of halos used for estimation.
Based on the results, we propose a unified
scheme to compare results from observational samples with theories. In the proposal, 
the scale at which bulk flow in a particular space of the Universe is measured is 
chosen to be the effective scale $R$ which is the radius of a spherical top-hat 
window function $W_{th}$ that yields the same bulk velocity dispersion $\sigma_V$
as the practical window function $W_O$ for the observational sample does in
linear theory. $W_O$ is not only determined by the sample geometry but also
contains weights emerged from selection function, incompleteness and etc., 
being analogous to the window function used in estimation of power spectrum.
Numerical experiments indicate that effects of selection functions on bulk flow
estimation could be corrected to a good accuracy by the simple treatment of Eq.~\ref{eq:sel}.

Variance ranges of bulk flow are clarified as well on the basis of Maxwellian
distribution in the work, we make a rough comparison of some recently measurements
with the $\Lambda$CDM model adopted in Pangu simulation, we find that
part results do deviate from the model by about $3\sigma$ but the tension between 
observation and model is not so strong as original works claimed.
Estimated effective scales for observation results in Figure~\ref{fig:obs} are
in fact the upper limits, the true effective scales could be even smaller
since we have assumed that those samples are of full sky coverage and their
selection functions have been corrected for during estimation. Furthermore,
observed bulk velocity consists of residuals from thermal motion of galaxies in their
host halo, which is not included in calculation of the velocity
dispersion so far. More accurate modeling could be developed
by assuming velocities of galaxies relative to their halos obey 
certain simple distribution, but it
requires explicit knowledge of occupation details of galaxies in host halo,
which in itself is already a challenging problem. A better way would be to
deduct the random motion component in the estimation procedure, 
such as the treatment in \citet{Wang2007}.

Correlation between bulk flow and dipole of internal mass is very weak, but
is stronger for larger volume. If one happens to be living in
a volume of radius greater than $100h^{-1}$Mpc with large bulk velocity, 
their observed mass dipole in the volume will
have considerable chance of being unusually strong. Typical misalignment angle between
bulk flow and mass dipole is mostly likely around $\sim 30-50^\circ$. This might
introduce non-negligible systematical bias to cosmological probes involving
local mass distribution, such as the late-time integrated 
Sachs-Wolfe effect \citep{ReesSciama1968}.

In our simulation there do exist volume of scale extending to
$200h^{-1}$Mpc in diameter moving with extreme large bulk velocity more than
$500\kms$. Most halos inside the volume are moving in alignment with the bulk flow
within $60^\circ$, and the flow shows no dependence on halo mass. Such group motion of
numerous halos will generate prominent kSZ signals, 
it is a rare event, but given its high speed and huge scale ($200h^{-1}$Mpc versus
$1h^{-1}$Gpc), probability of detection
is actually not very small, which of course also depends on the inclination between
bulk flow and line-of-sight. Another possible observation effect is that galaxies
in the volume could be dimmed or brightened on average by the
extreme bulk flow than galaxies in other places, 
which is the starting point of the effort tried by
\citet{NusserEtal2011} and \citet{AbateFeldman2012}. 

Dipoles of mass outside the
largest bulk flow region as environment are not tightly aligned with the bulk 
velocity, but deviate from it by around $\sim 20-40^\circ$. Interestingly we identified
that halo clustering of the particular volume is strengthened apparently
at scales $r\gtrsim 20h^{-1}$Mpc, simulation results point out that such enhancement 
is not completely accidental, at large scales two-point correlation function
of halos in a finite volume is indeed mildly correlated with bulk velocity. 
Bulk velocity is dominated by Fourier modes of velocity at scales larger than
the characteristic scale of the sample while in linear theory 
$\vect{v}(\vect{k}) \propto i\vect{k}\delta(\vect{k})/k^2$, unusually large
bulk velocity seems to imply that there should be extraordinary super large 
mode of density fluctuation topping up in the region. However bulk flow is hardly 
correlated with the total number or mass of enclosed halos inside the sample volume,
and as we checked the total number or mass of halos in the cell with largest 
bulk velocity is less than the mean value but still within $2\sigma$ variance 
range. Moreover, in linear regime Fourier modes of density fluctuation are
independent, $\xi$ of halos inside the volume is controlled actually by modes of scale
less than the characteristic scale of the sample. 

Aside from the theoretical puzzle, one question is whether the power excess
of halo clustering in a region at scales $\gtrsim 20h^{-1}$Mpc can be used as 
indicator of candidate space of extremely large bulk flow, the advantage of 
using two-point correlation function is that clustering does not
rely on direction of line-of-sight, the complication resulted from
redshift distortion in principle can be overcame by the ratio of
$\xi_h(r\sim 20-50h^{-1}{\rm Mpc})$ to $\xi_h$ at small scales e.g. $\sim 10h^{-1}$Mpc
where $\xi_h$ is not correlated with bulk flow. A more serious
concern is that if we are unluckily (or luckily) in a special region as large
as our current largest galaxy survey with extremely large
bulk flow, the measured clustering strength at and beyond scale of baryonic acoustic
oscillation would be significantly leveled up, could it be the case of power
excess at very large scales in the Baryon Oscillation Spectroscopic Survey (BOSS) 
elaborated by \citet{RossEtal2012}? 
To answer all these queries one surely needs multiple realizations of simulation 
of volume much bigger than our Pangu simulation, for the moment in this paper we 
have to leave these questions open.

\section*{Acknowledgment}
This work is partly supported by the NSFC through grants of 
Nos. 10873027, 10873035, 11073055, and 11133003.
YPJ, WPL, XHY and PJZ are members of the Innovation group funded by NSFC (No. 11121062).
JP and XK acknowledge the One-Hundred-Talent fellowships of CAS.
We appreciate stimulating discussion with Jiasheng Huang, Cheng Li, Guoliang Li and Lifan Wang,
as well as the very helpful comments and suggestion of the anonymous referee.

The Pangu simulation was carried out in the 
Supercomputing center of CNIC, CAS,
under the collaboration scheme of the Computational Cosmology Consortium of China (C4),
participating institutions are NAOC, PMO, SHAO and CNIC.


\begin{thebibliography}{48}
\expandafter\ifx\csname natexlab\endcsname\relax\def\natexlab#1{#1}\fi

\bibitem[{{Abate} \& {Feldman}(2012)}]{AbateFeldman2012}
{Abate}, A., \& {Feldman}, H.~A. 2012, \mnras, 419, 3482

\bibitem[{{Afshordi} {et~al.}(2009){Afshordi}, {Geshnizjani}, \&
  {Khoury}}]{AfshordiEtal2009}
{Afshordi}, N., {Geshnizjani}, G., \& {Khoury}, J. 2009, \jcap, 8, 30

\bibitem[{{Amanullah} {et~al.}(2010)}]{AmanullahEtal2010}
{Amanullah}, R., {et~al.} 2010, \apj, 716, 712

\bibitem[{{Bahcall} {et~al.}(1994){Bahcall}, {Cen}, \&
  {Gramann}}]{BahcallEtal1994}
{Bahcall}, N.~A., {Cen}, R., \& {Gramann}, M. 1994, \apjl, 430, L13

\bibitem[{{Ciecielg} \& {Chodorowski}(2004)}]{CiecielgChodorowski2004}
{Ciecielg}, P., \& {Chodorowski}, M.~J. 2004, \mnras, 349, 945

\bibitem[{{Colin} {et~al.}(2011){Colin}, {Mohayaee}, {Sarkar}, \&
  {Shafieloo}}]{ColinEtal2011}
{Colin}, J., {Mohayaee}, R., {Sarkar}, S., \& {Shafieloo}, A. 2011, \mnras,
  414, 264

\bibitem[{{Courtois} {et~al.}(2011){Courtois}, {Tully}, {Makarov}, {Mitronova},
  {Koribalski}, {Karachentsev}, \& {Fisher}}]{CourtoisEtal2011}
{Courtois}, H.~M., {Tully}, R.~B., {Makarov}, D.~I., {Mitronova}, S.,
  {Koribalski}, B., {Karachentsev}, I.~D., \& {Fisher}, J.~R. 2011, \mnras,
  414, 2005

\bibitem[{{Dai} {et~al.}(2011){Dai}, {Kinney}, \& {Stojkovic}}]{DaiEtal2011}
{Dai}, D.-C., {Kinney}, W.~H., \& {Stojkovic}, D. 2011, \jcap, 4, 15

\bibitem[{{Feldman} {et~al.}(2010){Feldman}, {Watkins}, \&
  {Hudson}}]{FeldmanEtal2010}
{Feldman}, H.~A., {Watkins}, R., \& {Hudson}, M.~J. 2010, \mnras, 407, 2328

\bibitem[{{G{\'o}rski} {et~al.}(2005){G{\'o}rski}, {Hivon}, {Banday},
  {Wandelt}, {Hansen}, {Reinecke}, \& {Bartelmann}}]{GorskiEtal2005}
{G{\'o}rski}, K.~M., {Hivon}, E., {Banday}, A.~J., {Wandelt}, B.~D., {Hansen},
  F.~K., {Reinecke}, M., \& {Bartelmann}, M. 2005, \apj, 622, 759

\bibitem[{{Haugb{\o}lle} {et~al.}(2007){Haugb{\o}lle}, {Hannestad}, {Thomsen},
  {Fynbo}, {Sollerman}, \& {Jha}}]{HaugbolleEtal2007}
{Haugb{\o}lle}, T., {Hannestad}, S., {Thomsen}, B., {Fynbo}, J., {Sollerman},
  J., \& {Jha}, S. 2007, \apj, 661, 650

\bibitem[{{Jarosik} {et~al.}(2011)}]{JarosikEtal2011}
{Jarosik}, N., {et~al.} 2011, \apjs, 192, 14

\bibitem[{{Jha} {et~al.}(2007){Jha}, {Riess}, \& {Kirshner}}]{JhaEtal2007}
{Jha}, S., {Riess}, A.~G., \& {Kirshner}, R.~P. 2007, \apj, 659, 122

\bibitem[{{Kashlinsky} {et~al.}(2011){Kashlinsky}, {Atrio-Barandela}, \&
  {Ebeling}}]{KashlinskyEtal2011}
{Kashlinsky}, A., {Atrio-Barandela}, F., \& {Ebeling}, H. 2011, \apj, 732, 1

\bibitem[{{Kashlinsky} {et~al.}(2010){Kashlinsky}, {Atrio-Barandela},
  {Ebeling}, {Edge}, \& {Kocevski}}]{KashlinskyEtal2010}
{Kashlinsky}, A., {Atrio-Barandela}, F., {Ebeling}, H., {Edge}, A., \&
  {Kocevski}, D. 2010, \apjl, 712, L81

\bibitem[{{Lahav} {et~al.}(1991){Lahav}, {Lilje}, {Primack}, \&
  {Rees}}]{LahavEtal1991}
{Lahav}, O., {Lilje}, P.~B., {Primack}, J.~R., \& {Rees}, M.~J. 1991, \mnras,
  251, 128

\bibitem[{{Landy} \& {Szalay}(1993)}]{LandySzalay1993}
{Landy}, S.~D., \& {Szalay}, A.~S. 1993, \apj, 412, 64

\bibitem[{{Lavaux} {et~al.}(2012){Lavaux}, {Afshordi}, \&
  {Hudson}}]{LavauxEtal2012}
{Lavaux}, G., {Afshordi}, N., \& {Hudson}, M.~J. 2012, ArXiv e-prints,
  astro-ph.CO/1207.1721

\bibitem[{{Lewis} {et~al.}(2000){Lewis}, {Challinor}, \&
  {Lasenby}}]{LewisEtal2000}
{Lewis}, A., {Challinor}, A., \& {Lasenby}, A. 2000, \apj, 538, 473

\bibitem[{{Linder}(2005)}]{Linder2005}
{Linder}, E.~V. 2005, \prd, 72, 043529

\bibitem[{{Ma} \& {Scott}(2012)}]{MaScott2012}
{Ma}, Y.-Z., \& {Scott}, D. 2012, ArXiv e-prints, astro-ph.CO/1208.2028

\bibitem[{{Macaulay} {et~al.}(2011){Macaulay}, {Feldman}, {Ferreira}, {Hudson},
  \& {Watkins}}]{MacaulayEtal2011}
{Macaulay}, E., {Feldman}, H., {Ferreira}, P.~G., {Hudson}, M.~J., \&
  {Watkins}, R. 2011, \mnras, 414, 621

\bibitem[{{Mak} {et~al.}(2011){Mak}, {Pierpaoli}, \& {Osborne}}]{MakEtal2011}
{Mak}, D.~S.~Y., {Pierpaoli}, E., \& {Osborne}, S.~J. 2011, \apj, 736, 116

\bibitem[{{Mersini-Houghton} \& {Holman}(2009)}]{MersiniHolman2009}
{Mersini-Houghton}, L., \& {Holman}, R. 2009, \jcap, 2, 6

\bibitem[{{Moscardini} {et~al.}(1996){Moscardini}, {Branchini}, {Brunozzi},
  {Borgani}, {Plionis}, \& {Coles}}]{MoscardiniEtal1996}
{Moscardini}, L., {Branchini}, E., {Brunozzi}, P.~T., {Borgani}, S., {Plionis},
  M., \& {Coles}, P. 1996, \mnras, 282, 384

\bibitem[{{Nusser} {et~al.}(2011){Nusser}, {Branchini}, \&
  {Davis}}]{NusserEtal2011}
{Nusser}, A., {Branchini}, E., \& {Davis}, M. 2011, \apj, 735, 77

\bibitem[{{Nusser} \& {Davis}(1994)}]{NusserDavis1994}
{Nusser}, A., \& {Davis}, M. 1994, \apjl, 421, L1

\bibitem[{{Nusser} \& {Davis}(2011)}]{NusserDavis2011}
---. 2011, \apj, 736, 93

\bibitem[{{Nusser} {et~al.}(1991){Nusser}, {Dekel}, {Bertschinger}, \&
  {Blumenthal}}]{NusserEtal1991}
{Nusser}, A., {Dekel}, A., {Bertschinger}, E., \& {Blumenthal}, G.~R. 1991,
  \apj, 379, 6

\bibitem[{{Osborne} {et~al.}(2011){Osborne}, {Mak}, {Church}, \&
  {Pierpaoli}}]{OsborneEtal2011}
{Osborne}, S.~J., {Mak}, D.~S.~Y., {Church}, S.~E., \& {Pierpaoli}, E. 2011,
  \apj, 737, 98

\bibitem[{{Park} \& {Park}(2006)}]{ParkPark2006}
{Park}, C.-G., \& {Park}, C. 2006, \apj, 637, 1

\bibitem[{Pham-Gia {et~al.}(2006)Pham-Gia, Turkkan, \&
  Marchand}]{Pham-GiaEtal2007}
Pham-Gia, T., Turkkan, N., \& Marchand, E. 2006, Communications in Statistics -
  Theory and Methods, 35, 1569

\bibitem[{{Rees} \& {Sciama}(1968)}]{ReesSciama1968}
{Rees}, M.~J., \& {Sciama}, D.~W. 1968, \nat, 217, 511

\bibitem[{{Ross} {et~al.}(2012)}]{RossEtal2012}
{Ross}, A.~J., {et~al.} 2012, \mnras, 424, 564

\bibitem[{{Sandage} {et~al.}(2010){Sandage}, {Reindl}, \&
  {Tammann}}]{SandageEtal2010}
{Sandage}, A., {Reindl}, B., \& {Tammann}, G.~A. 2010, \apj, 714, 1441

\bibitem[{{Saunders} {et~al.}(2000)}]{SaundersEtal2000}
{Saunders}, W., {et~al.} 2000, \mnras, 317, 55

\bibitem[{{Scoccimarro}(2004)}]{Scoccimarro2004}
{Scoccimarro}, R. 2004, \prd, 70, 083007

\bibitem[{{Song} {et~al.}(2011){Song}, {Sabiu}, {Kayo}, \&
  {Nichol}}]{SongEtal2011}
{Song}, Y.-S., {Sabiu}, C.~G., {Kayo}, I., \& {Nichol}, R.~C. 2011, \jcap, 5,
  20

\bibitem[{{Springel}(2005)}]{Springel2005}
{Springel}, V. 2005, \mnras, 364, 1105

\bibitem[{{Springob} {et~al.}(2007){Springob}, {Masters}, {Haynes},
  {Giovanelli}, \& {Marinoni}}]{SpringobEtal2007}
{Springob}, C.~M., {Masters}, K.~L., {Haynes}, M.~P., {Giovanelli}, R., \&
  {Marinoni}, C. 2007, \apjs, 172, 599

\bibitem[{{Strauss} \& {Willick}(1995)}]{StraussWillick1995}
{Strauss}, M.~A., \& {Willick}, J.~A. 1995, \physrep, 261, 271

\bibitem[{{Sunyaev} \& {Zeldovich}(1980)}]{SunyaevZeldovich1980}
{Sunyaev}, R.~A., \& {Zeldovich}, I.~B. 1980, \mnras, 190, 413

\bibitem[{{Szapudi}(1998)}]{Szapudi1998a}
{Szapudi}, I. 1998, \apj, 497, 16

\bibitem[{{Turnbull} {et~al.}(2012){Turnbull}, {Hudson}, {Feldman}, {Hicken},
  {Kirshner}, \& {Watkins}}]{TurnbullEtal2012}
{Turnbull}, S.~J., {Hudson}, M.~J., {Feldman}, H.~A., {Hicken}, M., {Kirshner},
  R.~P., \& {Watkins}, R. 2012, \mnras, 420, 447

\bibitem[{{Wang}(2007)}]{Wang2007}
{Wang}, L. 2007, ArXiv e-prints, astro-ph/0705.0363

\bibitem[{{Watkins} {et~al.}(2009){Watkins}, {Feldman}, \&
  {Hudson}}]{WatkinsEtal2009}
{Watkins}, R., {Feldman}, H.~A., \& {Hudson}, M.~J. 2009, \mnras, 392, 743

\bibitem[{{Weyant} {et~al.}(2011){Weyant}, {Wood-Vasey}, {Wasserman}, \&
  {Freeman}}]{WeyantEtal2011}
{Weyant}, A., {Wood-Vasey}, M., {Wasserman}, L., \& {Freeman}, P. 2011, \apj,
  732, 65

\bibitem[{{Wyman} \& {Khoury}(2010)}]{WymanKhoury2010}
{Wyman}, M., \& {Khoury}, J. 2010, \prd, 82, 044032

\end{thebibliography}

\end{document}